\documentclass[useAMS,usenatbib]{mn2e}

\usepackage{graphics}
\usepackage{psfig}
\usepackage{amsmath}
\usepackage{amssymb}
\usepackage{upgreek}
\usepackage{longtable}
\usepackage[mathscr]{eucal}
\usepackage[dvips]{graphicx}
\usepackage{longtable}
\usepackage{lscape}
\usepackage{float}

\title
[3D dynamics of young groups and clusters]
{The {\it Gaia}-ESO Survey: 3D dynamics of young groups and clusters from GES and {\it Gaia} EDR3}
\author
[Wright et al.]
{Nicholas J. Wright$^1$, R.D. Jeffries$^1$, R.J. Jackson$^1$, G.G. Sacco$^2$, Becky Arnold$^1$, 
\newauthor  E. Franciosini$^2$, G. Gilmore$^3$, A. Gonneau$^{3,4}$, L. Morbidelli$^2$, L. Prisinzano$^{5}$, 
\newauthor S. Randich$^2$, Clare C. Worley$^{3,6}$ \\
\\
$^{1}$Astrophysics Group, Keele University, Keele, ST5 5BG, UK\\
$^{2}$INAF – Osservatorio Astrofisico di Arcetri, Largo E. Fermi, 5, 50125 Firenze, Italy\\
$^{3}$Institute of Astronomy, University of Cambridge, Madingley Road, Cambridge CB3 0HA, UK\\
$^{4}$Université de Strasbourg, CNRS, Observatoire astronomique de Strasbourg, UMR 7550, F-67000 Strasbourg, France\\
$^{5}$INAF – Osservatorio Astronomico di Palermo, Piazza del Parlamento 1, 90134 Palermo, Italy\\
$^{6}$School of Physical and Chemical Sciences --- Te Kura Mat\={u}, University of Canterbury, Private Bag 4800, Christchurch 8140, New Zealand\\
}

\begin{document}
\maketitle

\begin{abstract}

We present the first large-scale 3D kinematic study of $\sim$2700 spectroscopically-confirmed young stars ($<$ 20 Myr) in 18 star clusters and OB associations (hereafter {\it groups}) from the combination of {\it Gaia} astrometry and {\it Gaia}-ESO Survey spectroscopy. We measure 3D velocity dispersions for all groups, which range from 0.61 to 7.4 km~s$^{-1}$ (1D velocity dispersions of 0.35 to 4.3 km~s$^{-1}$). We find the majority of groups have anisotropic velocity dispersions, suggesting they are not dynamically relaxed. From the 3D velocity dispersions, measured radii and estimates of total mass we estimate the virial state and find that all systems are super-virial when only the stellar mass is considered, but that some systems are sub-virial when the mass of the molecular cloud is taken into account. We observe an approximately linear correlation between the 3D velocity dispersion and the group mass, which would imply that the virial state of groups scales as the square root of the group mass. However, we do not observe a strong correlation between virial state and group mass. In agreement with their virial state we find that nearly all of the groups studied are in the process of expanding and that the expansion is anisotropic, implying that groups were not spherical prior to expansion. One group, Rho Oph, is found to be contracting and in a sub-virial state (when the mass of the surrounding molecular cloud is considered). This work provides a glimpse of the potential of the combination of {\it Gaia} and data from the next generation of spectroscopic surveys.

\end{abstract}


\section{Introduction}

Stars form in turbulent molecular clouds with a hierarchical and highly substructured spatial distribution that follows the distribution of dense gas \citep{elme02}. Many recently-formed stars are observed in groups or `clusters' of some sort, some showing substructure, while others are centrally concentrated with a smooth density distribution \citep{lada03,gute08}. The majority of groups quickly disperse, with very few young groups surviving as long-lived open clusters \citep{adam10}. The stellar groups that disperse are often briefly observed as `associations', gravitationally unbound and low-density groups of young stars \citep{wrig20,wrig22}.

Young star clusters may form monolithically within their parental molecular cloud \citep{bane15} or they may form hierarchically from mergers between sub-clusters \citep{bonn03,arno22}. Existing kinematic studies of young subgroups have found no evidence that they are in the process of merging \citep{kuhn19} and therefore if this process takes place if must happen very early, most likely while the subgroups are still heavily embedded. Directly observing these mergers may therefore be very difficult without infrared astrometry. However, kinematic signatures of the merger process could be observable after the merger \citep[e.g.,][]{park16,arno22}, including inverse energy equipartition, recently observed in the young cluster NGC~6530 \citep{wrig19b}, and hinting at the possibility of observing such signatures in other groups.

The evolution of a young group depends on its mass, structure, dynamics and the gravitational potential from the surrounding molecular gas \citep[e.g.,][]{park14}. Denser systems will usually evolve faster \citep{spit87}, mixing, relaxing, and evolving towards energy equipartition. The initial structure and state of the system, the distribution of stellar masses, and the external gravitational potential will all affect the timescale on which a given system evolves.

The vast majority of young groups do not survive to maturity as long-lived open clusters. The process of residual gas expulsion was long thought to be responsible for unbinding young clusters by dispersing the residual gas and its associated gravitational potential \citep{tutu78,lada84}. However, recent studies have questioned the role that residual gas expulsion may play \citep{dale15} or have considered other mechanisms for cluster dispersal such as tidal interactions \citep{krui12}. While some older studies of dispersing OB associations could not find clear evidence that they were dispersing from a more compact configuration \citep[e.g.,][]{wrig16,dzib17,wrig18,ward20}, recent studies that identify members of the association kinematically have found the majority show strong expansion patterns \citep{cant19b,kuhn19,arms20,quin21,quin23}.

The physical processes that drive the formation, evolution and dispersal of star clusters and associations are clearly poorly understood. This is due to the difficulty observing young stars that are often highly embedded in their parental molecular cloud, confirming the youth of stars in low-density OB associations, and diagnosing the highly-stochastic physical processes that are predominantly driven by gravitational attraction, a very difficult force to trace or directly observe. Kinematic information can allow the physical processes involved to be constrained, and this is quickly becoming more abundant thanks to {\it Gaia} and data from large-scale spectroscopic surveys such as the {\it Gaia}-ESO Survey (GES).

GES \citep{gilm22,rand22} is a large public survey programme that was carried out using the FLAMES \citep[Fibre Large Array Multi Element Spectrograph,][]{pasq02} fibre-fed spectrograph at the VLT (Very Large Telescope). The aim of the survey was to understand the formation and evolution of all components of our Galaxy, achieved by measuring radial velocities (RVs) and chemical abundances. Over a 6 year period the survey observed approximately $10^5$ stars in our galaxy, including in 62 young and open clusters, groups and associations \citep{rand22}.

In this paper we are targeting the 18 young ($< 20$ Myrs) clusters, associations and star-forming regions observed by the {\it Gaia}-ESO Survey, listed in Table~\ref{clusters}. In Appendix~A we provide a summary of the 18 groups studied with information from the literature. We combine GES spectroscopy and {\it Gaia} astrometry for thousands of stars towards these 18 young groups, confirming the youth of the stars using spectroscopy and combining spectroscopic RVs with astrometric proper motions (PMs) to facilitate the largest 3D kinematic study of young stellar groups to date. We chose to focus on young groups as their dynamics provide a valuable probe of the star cluster formation process, relatively unspoilt by dynamical mixing and evolution.

Section~2 introduces the targeted groups and their properties. Section~3 describes the spectroscopic and astrometric data used, while Section~4 outlines our process for confirming the youth of the targeted stars. In Section~5 we measure 3D velocity dispersions for all the groups, exploring a variety of models to determine how best to measure the velocity dispersion of a system, and then calculating the virial state of each system. In Section~6 we search for evidence of expansion or contraction in our sample using different methods to quantify the presence of expansion and the rate at which each group is expanding. In Section~7 we compare the kinematic properties of these groups with their physical properties and in Section~8 we discuss our results and their implications for the formation and dispersal of star clusters.

\section{Observational data}
\label{s_observations}

Here we describe the observational data, GES spectroscopy and {\it Gaia} early data release 3 (EDR3) astrometry for stars towards our 18 targeted groups. In the work that follows we discuss the 18 groups as separate targets, although in this section groups that were part of the same observing block were processed together before being separated into their constituent groups.

\begin{table*}
\caption{Young clusters and associations studied in this work in order of increasing distance. Ages and extinctions are taken from the literature, while distances are calculated as described in Section~\ref{s-background} (note that distance uncertainties combine the observational errors and fitting errors, but do not include any systematic errors, though the {\it Gaia} EDR3 parallax zero point of $-17$~$\mu$as has been corrected for). N$_{obs}$ is the number of stars with spectroscopy from GES, N$_{data}$ is the number of stars with either EW(Li) or FWZI(H$\alpha$) necessary to determine membership, N$_{yso}$ is the number of high-confidence young stars towards each group, N$_{1D+}$ is the number of objects with either a PM or a RV, and N$_{3D}$ is the number of young stars with 3D kinematics.}
\label{clusters} 
\begin{tabular}{lllllllllll}
\hline
Group			& Age	& Ref.	& Distance 				& $A_V$	& Ref.  & N$_{obs}$	& N$_{data}$ & N$_{yso}$ & N$_{1D+}$	& N$_{3D}$ \\
				& [Myr]	&		& [pc] 					& [mag.]	&		&		&		&		&	\\
\hline
Rho Ophiuchus			& 3.0		& GR21	& $136^{+1}_{-1}$		& 1.0		& W08	& 310	& 281	& 38		& 38		& 30 \\
Cha I (South)			& 1.5		& GA21	& $187^{+1}_{-1}$		& 3.0		& L07	& 345	& 305	& 48		& 48		& 28	\\
Cha I (North)			& 1.5		& GA21	& $191^{+1}_{-1}$		& 3.0		& L07	& 362	& 334	& 45		& 45		& 26 \\
Gamma Vel			& 19.5	& J17	& $334^{+1}_{-1}$		& 0.13	& J09	& 124	& 119	& 97		& 97		& 95 \\
Vela OB2				& 14		& A22	& $367^{+2}_{-2}$		& 0.13	& J09	& 989	& 916	& 73		& 73		& 72 \\
25 Ori				& 19		& F22	& $339^{+1}_{-1}$		& 0.3		& Z19	& 245	& 227	& 149	& 149	& 115 \\
Barnard 30			& 2.4		& K18	& $384^{+2}_{-2}$		& 0.4		& Z19	& 227	& 196	& 57		& 57		& 36 \\
$\lambda$ Ori			& 10		& B13	& $389^{+1}_{-1}$		& 0.4		& Z19	& 380	& 329	& 144	& 144	& 111 \\
Barnard 35			& 2.6		& K18	& $390^{+1}_{-1}$		& 0.4		& Z19	& 226	& 183	& 57		& 57		& 39 \\
S Mon Cluster (NGC 2264) & 2 	& V19	& $690^{+2}_{-2}$		& 0.71	& D08	& 863	& 780	& 246	& 244	& 192 \\
Spokes Cluster (NGC 2264) & 1	& V19	& $691^{+2}_{-2}$		& 0.71	& D08	& 992	& 881	& 281	& 279	& 188 \\
ASCC 50 / RCW33		& 5		& This work & $912^{+3}_{-3}$		& 0.71	& K13	& 1224	& 1131	& 193	& 190	& 155 \\
Collinder 197			& 5		& B10	& $925^{+7}_{-7}$		& 1.05	& B10	& 408	& 387	& 140	& 140	& 110 \\
NGC 6530			& 1.5		& B13	& $1242^{+33}_{-32}$	& 1.1		& S00	& 1982	& 1687	& 644	& 635	& 455 \\
NGC 2244			& 2		& B13	& $1368^{+24}_{-23}$	& 1.5		& M19	& 248	& 216	& 109	& 109	& 88 \\
NGC 2237			& 2		& W10	& $1494^{+44}_{-41}$	& 1.5		& M19	& 54		& 54		& 23		& 23		& 20 \\
Trumpler 14			& 2.5		& H12, D17 & $2211^{+16}_{-16}$	& 2.0		& H12	& 631	& 363	& 162	& 152	& 82 \\
Trumpler 16			& 2		& H12, D21 & $2260^{+23}_{-22}$	& 2.0		& H12	& 1269	& 616	& 208	& 202	& 100 \\
\hline
Totals			&		&		&						&		&		& 10,879	& 9,005	& 2,683	& 2,651	& 1,914 \\
\hline
\end{tabular}
\flushleft References: A22 \citep{arms22}, B10 \citep{bona10}, B13 \citep{bell13}, D08 \citep{dahm08}, D17 \citep{dami17}, D21 \citep{dias21}, F22 \citep{fran22}, GA21 \citep{gall21}, GR21 \citep{gras21}, H12 \citep{hur12}, J09 \citep{jeff09}, J17 \citep{jeff17}, K13 \citep{khar13}, K18 \citep{koun18}, L07 \citep{luhm07}, M19 \citep{muzi19}, S00 \citep{sung00}, V19 \citep{venu19}, W08 \citep{wilk08}, W10 \citep{wang10}, Z19 \citep{zari19}.
\end{table*}

\subsection{{\it Gaia}-ESO Survey spectroscopy}

The main source of data for this work is spectroscopy from the sixth and final internal data release (iDR6) of GES. GES targets are selected using homogeneous criteria based on the positions of stars in various colour magnitude diagrams (CMD), unbiased with respect to kinematics, though the exact strategy varies from group to group \citep{brag22}. The observations of each group are not complete within any area or magnitude range, being limited by the field of view of the instrument and the difficulty placing fibres close to each other. This can introduce slight biases against targets in dense areas or on the periphery of groups, though thanks to the large number of observing blocks and the good spatial coverage of the blocks, these biases are not significant.

GES data were downloaded from the Edinburgh Wide Field Astronomy Unit\footnote{http://ges.roe.ac.uk}. GES data are reduced and analysed using common methods and software to produce a uniform set of spectra and extracted stellar parameters. These methods are described in \citet{jeff14} and \citet{sacc14} for the data reduction, and in \citet{lanz15}, \citet{panc17} and \citet{hour23} for the data analysis, calibration and extraction of stellar parameters of stars in young groups. The final full GES data set is also now available through the ESO archive. 

In total, spectroscopy for 11,247 targets in our 11 observational datasets was used. These were then subdivided into 18 groups (see Section~\ref{s_targets} for more details) and trimmed of spatial non-members (e.g., in outlying fields not associated with the targeted group), which lead to 10,897 targets, as listed in Table~\ref{clusters}.

Where possible, stellar effective temperatures derived directly from spectroscopy were used (available for 8871 or 79\% of sources). Where these were not available we estimated temperatures using the $\tau$ spectral index introduced by \citet{dami14} and calculated for most GES spectra (available for an extra 767 or 7\% of sources). We calibrated a new relationship between $T_{eff}$ and $\tau$ for young stars ($<$ 50 Myr) using sources with both spectroscopic effective temperatures and $\tau$ values. Uncertainties were estimated based on the spread in temperature at a given $\tau$ value. For sources that lacked a spectroscopic temperature and a $\tau$ index we estimated effective temperatures from the available photometry, dereddened using the average group extinction (Table~\ref{clusters}), extinction coefficients from \citet{schl98} and \citet{dani18}, and the tables of intrinsic colour from \citet{peca16}, prioritising the colour with the longest baseline. Uncertainties were estimated from the photometric uncertainty and the spread in temperature at a given colour. This was possible for a further 1569 or 14\% of sources. For the remaining 38 (0.3\%) of sources this was impossible either due to a lack of photometry or the available photometry being inconsistent with group membership.

RVs are determined using a combination of a cross-correlation method and a direct modelling approach \citep{gilm22}. For the RV uncertainties we use the improved empirical precision provided by \citet{jack15}, which is calculated from a combination of the target signal-to-noise ratio, the projected equatorial velocity ($v$ sin $i$) and empirically-derived constants \citep[see][for the values and derivation of these]{jack20}.

Lithium equivalent widths and H$\alpha$ full widths at half maximum are determined using direct profile integration, taking into account the star's RV, $v$~sin~$i$ and S/N \citep[see][for more details]{fran22b}. Values of the gravity-sensitive spectral index $\gamma$ are defined and calibrated according to \citet{dami14}. All of these values are available in the GES data set at the ESO archive.

Table~\ref{clusters} lists the groups targeted in this study, including the number of targets that were observed with GES towards each group. In total we have spectroscopy for 11,247 objects, of which 10,897 are towards our 18 groups, and of these 10,859 (99.7\%) have effective temperatures, 8668 (80\%) have RVs, 7366 (68\%) have lithium equivalent widths, and 9321 (86\%) have $\gamma$ indices.

\subsection{{\it Gaia} EDR3 astrometry and photometry}
\label{s-gaia_dr3}

We use astrometry and photometry from {\it Gaia} \citep{prus16} EDR3 \citep{brow20}, which contains parallaxes, $\varpi$, and PMs, $\mu_\alpha$ and $\mu_\delta$, calculated from the first 34 months of {\it Gaia} observations. These data achieve an extremely high level of astrometric precision for a sample of unprecedented size, though it is still calculated assuming single-star behaviour \citep{lind21}. EDR3 is effectively complete in the magnitude range $G = 3 - 18.5$~mag, with a detection threshold of $G = 20.7$~mag. Extremely bright ($G < 3$~mag) or high PM ($> 0.6$~arcsec/yr) stars, or objects in very crowded areas of the sky (predominantly in globular clusters), can suffer from incompleteness \citep{fabr21}, though this should not significantly affect this work. The systematic uncertainties in parallax and PM that were present in DR2 have been greatly reduced, with less variation over the sky \citep{brow20,lind21}. Comparison with quasars and known binaries indicates that the parallax zero-point has been reduced compared to DR2 and is now $\sim$ -17~$\mu$as, a correction applied to all parallaxes when calculating distances. The effect of this correction is largest for the most distant group, Trumpler~16, where it results in a reduction in the inferred distance of $\sim$4\%. The affect is smaller for all other groups and $<$2\% for groups within 1~kpc.

GES sources were cross-matched with {\it Gaia} EDR3 using a radius of 1$^{\prime\prime}$, which resulted in a total of 11,179 matches (99.4\%). {\it Gaia} astrometry and photometry was then filtered based on the criteria outlined in the {\it Gaia} data release papers and technical notes. For the astrometry we required the `re-normalised unit weighted error' (RUWE) to be less than or equal to 1.4 \citep{lind21}. This removes sources with spurious astrometry and helps filter contamination from double stars, astrometric effects from binary stars and other contamination problems\footnote{Note that we do not discard the GES RVs for stars failing the RUWE cut, even though they might be binaries, because we do not know the fraction of binary stars that such a cut would discard and would prefer to model the effects of binarity on the velocity dispersion (see Section~\ref{s-forwardmodel}) rather than exclude them.}. Applying this cut removed astrometry for 1407 sources (12.6\%), leaving 9772 sources with {\it Gaia} EDR3 astrometry (sources without valid astrometry were not completely discarded as they might still have useful RVs from GES). {\it Gaia} photometry is believed to be well-behaved for sources with $G < 18$, as is the case for the vast majority of our sources, and so no filtering of the photometry was performed.

We do not use stellar effective temperatures from {\it Gaia} EDR3 since they are derived under the assumption of no reddening, which is invalid for our targets. We also don't use {\it Gaia} RVs for our sources as they are limited to the brightest sources and generally inferior to the GES data.

In the astrometric analysis that follows we always use the full ($5 \times 5$) {\it Gaia} covariance matrix whenever propagating uncertainties, and consider all errors to follow a normal distribution \citep{aren18}.

\section{Membership selection}

In this section we outline how we have identified the young stars in each group studied. GES spectroscopy is not complete, either spatially or in any magnitude range, and therefore our approach to membership selection is not to attempt to be complete, but to derive a reliable list of members that is as free from contamination as possible. To avoid any kinematic biases we have not applied any sort of membership criteria based on RVs or PMs.

The main source of contamination for candidate young stars selected from a CMD (as GES targets were) are foreground main sequence stars and background giants. Distinguishing young stars from foreground main sequence stars requires an indicator of youth such as the presence of photospheric lithium or strong H$\alpha$ emission. Identifying and removing background giants can be achieved using either a spectroscopic surface gravity indicator to discern giants from pre-main sequence stars or by using parallaxes.

GES spectroscopy provides three separate measures that can be used for this purpose: the equivalent width of the lithium 6708~\AA\ line, the gravity index $\gamma$ \citep{dami14}, and the full width at zero intensity (FWZI) of the H$\alpha$ line. These three parameters have been used by multiple GES studies to define samples of young stars free of kinematic bias \citep[e.g.,][]{jeff14,rigl16,sacc17,brav18,wrig19}, and we follow the same approach here, complementing this with the use of {\it Gaia} parallaxes.

\subsection{Separating foreground main-sequence stars}

Main sequence stars of certain types can be separated from young, late-type stars based on the presence of lithium in their atmospheres. Late-type stars deplete their primordial lithium after approximately 30 (for M-type stars) to 300~Myr (for K-type stars) due to burning and subsequent mixing throughout the convection zone \citep{sode10}. The presence of lithium in the atmosphere of K-M stars is therefore an effective indicator of youth. We follow the method of \citet{wrig19} and require our targets to have EW(Li) values greater than observed in stars of the same temperature in the 30--50~Myr cluster IC~2602 \citep{rand97}. Out of our 10,897  targets, 7366 have measures of EW(Li) (excluding upper limits) and of these 2384 pass our membership criteria (32.4\%).

We note that the EW(Li) can be underestimated in stars with high accretion rates if the continuum emission in excess produced by the accretion shock reduces the measured signal \citep[e.g.,][]{pall05}. From the 8513 stars that fail the EW(Li) membership test (or lack equivalent widths to perform such a test), 2659 have FWZI(H$\alpha$) measurements, and 1096 of these (41.2\%) have FWZI(H$\alpha$) greater than 4~\AA\ \citep{boni13,pris19} and are therefore re-classified as members. This leaves 3480 stars as likely young stars based on their spectroscopic properties. 

Figures~\ref{spectral_membership1} and \ref{spectral_membership2} show both EW(Li) and FWZI(H$\alpha$) as a function of $T_{eff}$ for all stars in the 18 groups studied.

\subsection{Separating background giants}
\label{s-background}

To separate young stars from background giants (which sometimes show lithium in their atmospheres) we use the surface gravity-sensitive index $\gamma$ \citep{dami14}, which is capable of distinguishing between main sequence stars, pre-main sequence stars and giants \citep[cool giants with $T_{eff} < 5600$~K have $\gamma > 1$,][]{dami14}. Of our 3480 current candidate members, 3326 have $\gamma$ measurements, and of these 413 (12.4\%) have $\gamma$ values that suggest they could be background giants and are therefore removed from our sample. This leaves 2913 stars that we retain as likely young stars. We also retained as likely young stars the 154 stars that lack $\gamma$ measurements, since the probability of these being giants is relatively low ($\sim$12\%) and will be removed by the next step. Figures~\ref{spectral_membership1} and \ref{spectral_membership2} show $\gamma$ as a function of $T_{eff}$ for the stars in all 18 groups studied.

As a further check to remove non-members we implement a parallax cut using data from {\it Gaia} EDR3, which is available for 2532 of the 3067 likely young stars. For each group we fit the parallax distribution with a single Gaussian and derive central values and dispersions. The dispersions of the parallax distributions are corrected for the contribution of the nonuniform parallax errors using the method of \citet{ivez14} to calculate parallax dispersions for each group. We then exclude all sources with parallaxes more than two standard deviations from the central value, where the standard deviation is the combination of the intrinsic parallax dispersion for the group and the measurement uncertainty for each star. A cut at 2$\sigma$ was chosen to balance the need to include the majority of members whilst rejecting a reasonable number of non-members. The Col. 197 group was found to be projected against another group of young stars in the background with parallaxes of $\varpi = 0.5$--$0.9$ mas (see Figure~\ref{spectral_membership2}) that we isolated by ensuring that the parallax distribution was fitted to the main peak at $\varpi \sim 1.1$~mas that represents Col. 197.

This cut excludes 384 (15.2\%) of the 2532 likely young stars with parallaxes, leaving 2148 high-confidence young stars. We retained the 535 stars without parallaxes as the probability of these being non-members is low ($\sim$14\%) and many of the background giants were already removed by the previous step. Note that we do not use parallax to add in any new stars, only to remove background contaminants. Figures~\ref{spectral_membership1} and \ref{spectral_membership2} show both $\gamma$ as a function of $T_{eff}$ and the parallax distribution for all candidate members.

\subsection{Final sample of members}

Our final sample contains 2683 spectroscopically-confirmed young stars distributed between the 18 groups, of which 1914 have 3D kinematics (71\%), 503 only have RVs (18.7\%), 234 only have PMs (8.7\%), and 32 lack both PMs and RVs (1.2\%), the latter of which are discarded to leave a sample of 2651 young stars with either RVs, PMs or both. We do not require the stars in our sample to have both RVs and PMs so as to maximise the numbers of stars with reliable kinematic data in at least one dimension that we can use to constrain the kinematic properties of the groups and clusters studied. Following the Bayesian approach used in this paper there should be no downsides to this, and we have tested this with additional simulations and can confirm that no systematic biases arise (see Section~\ref{s-forwardmodel}).

The median RV uncertainty for stars in our sample is 0.59~km~s$^{-1}$, with some as low as 0.25~km~s$^{-1}$. The median PM uncertainty is 0.062 mas~yr$^{-1}$, which equates to between 0.04 and 0.80 km~s$^{-1}$ depending on the distance to the target group. The median parallax uncertainty is 0.074~mas.

Our full catalogue of members, including their photometry, astrometry, and spectroscopic parameters is included in an online table made available on VizieR.

\section{3D velocity distributions and dispersions}

\begin{figure*}
\centering
\includegraphics[width=440pt,trim=50 100 50 260]{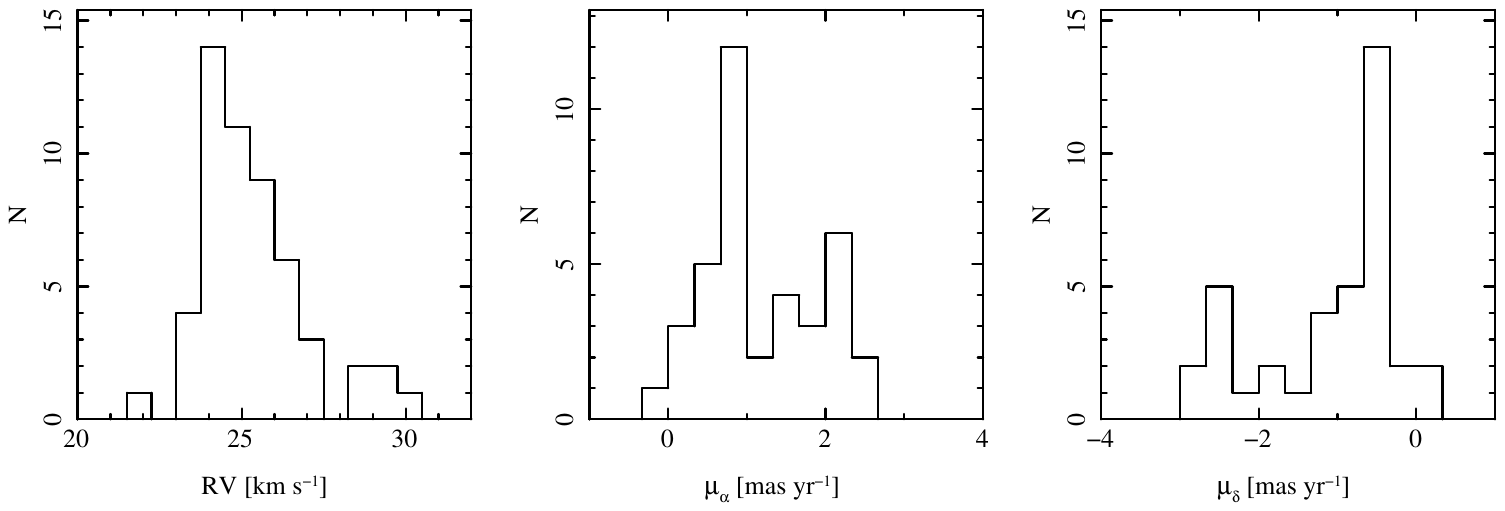}
\caption{Velocity distributions for young stars in Barnard 30 (53 stars with RVs, 38 stars with PMs).}
\label{example_velocity_dispersions}
\end{figure*}

\begin{figure*}
\centering
\includegraphics[width=500pt,trim=0 0 0 0]{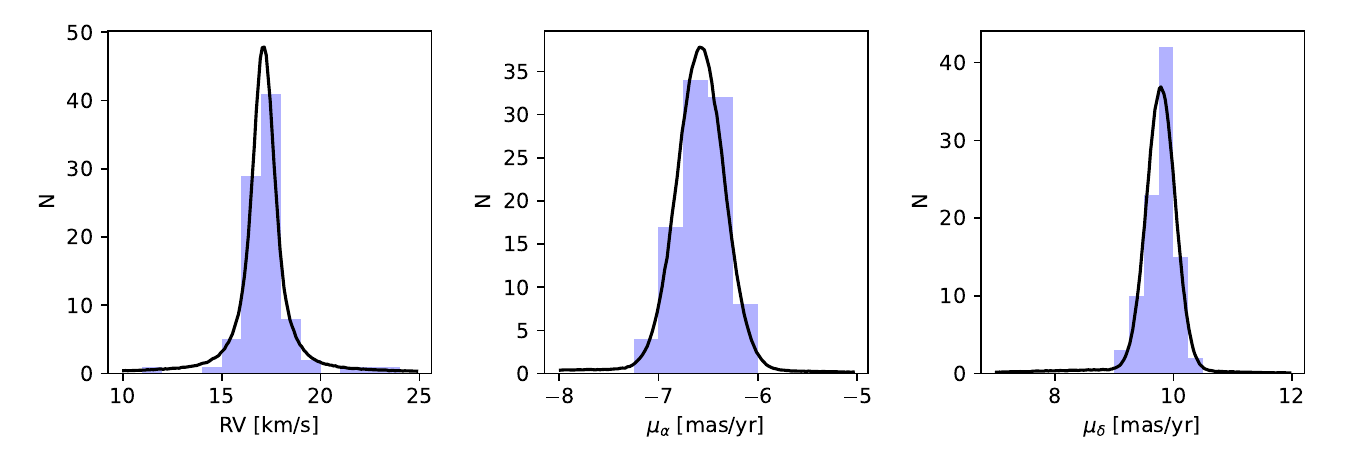}
\caption{Velocity distributions (blue histograms) for young stars in the Gamma Velorum cluster (97 stars with RVs, 95 stars with PMs) with our forward-modelled, non-rotated velocity dispersion fits superimposed (black lines).}
\label{velocity_dispersion_fit_nonrotated}
\end{figure*}

\begin{figure*}
\centering
\includegraphics[width=400pt,trim=0 0 0 0]{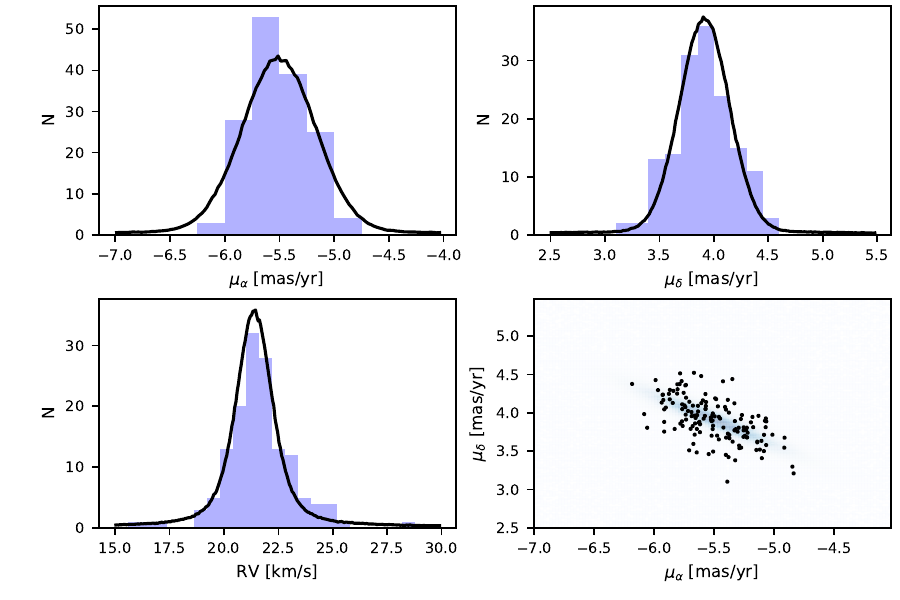}
\caption{Velocity distributions for young stars in ASSC~50 (190 stars with RVs, 155 stars with PMs) with our forward-modelled, partially-rotated velocity dispersion fits superimposed. The top panels show the two PM velocity distributions (blue histograms) with projected velocity dispersion fits (black lines), the bottom left panel shows the RV velocity distribution (blue histogram) with the velocity dispersion fit (black line), and the bottom right panel shows a 2D projection of the PM velocity distribution (black points) with the projected velocity dispersion fit as a blue 2D histogram.}
\label{velocity_dispersion_fit_2Drotated}
\end{figure*}

Here we present and analyse the 3D kinematics of stars in the young groups studied. We start by considering the 3D velocity distributions for each group, calculating velocity dispersions in all three dimensions for each group using a forward-modelling Bayesian approach, and from this derive virial masses for each group that allows us to assess their gravitational boundedness.

\subsection{Velocity distributions}
\label{sec:distributions}

The velocity distributions of stars in the 18 groups studied show a variety of morphologies. A small number of the velocity distributions are broadly Gaussian, albeit with extended wings that might constitute runaway stars, binaries or non-members of the group within our fields of view. However many of the velocity distributions deviate from normality, sometimes subtly and other times with very clear kinematic substructure that is evident even in 1D velocity distributions. An example is shown in Figure~\ref{example_velocity_dispersions} for young stars in Barnard 30. The velocity distributions in all three dimensions show evidence for substructure, with clear bimodal distributions in $\mu_\alpha$ and $\mu_\delta$. We estimate that of the 18 groups, 8 show evidence for kinematic substructure in at least one dimension.

We conducted Shapiro-Wilk tests of normality on the velocity distributions of stars in all 3 dimensions of the 18 groups. All of the groups except five show highly significant ($>$3$\sigma$) evidence for non-Gaussianity in all three dimensions (significance of rejecting the null hypothesis of a normal distribution). The groups whose velocity distributions are consistent with Gaussianity in at least one dimension are primarily the poorly-sampled groups (Cha I South, Rho Oph, Barnard 30 and Barnard 35) and therefore this may just be that there are insufficient stars to confidently rule out a normal distribution. The Gamma Vel cluster is the only well-sampled region ($\sim$100 stars) whose velocity distribution is not significantly non-Gaussian in more than one dimension ($\mu_\alpha$ and $\mu_\delta$), most likely because the cluster is relatively old ($\sim$19.5 Myr) and therefore more dynamically processed.

\subsection{Velocity dispersion fitting}
\label{s-forwardmodel}

To calculate the velocity dispersion for these groups we use Bayesian inference and model the distribution of observed velocities using a simple parameterised model that we compare with the observations in a probabilistic way \citep[see e.g.,][]{wrig19}. The aim is to determine which of the various sets of parameters, $\boldsymbol{\theta}$, best explain the observations, $\boldsymbol{d}$. In Bayes's theorem this is known as the posterior distribution, $\boldsymbol{P(\theta | d)} = \boldsymbol{P(d | \theta)} \, \boldsymbol{P(\theta)} / \boldsymbol{P(d)}$, where $\boldsymbol{P(d | \theta)}$ is the likelihood model, $\boldsymbol{P(\theta)}$ are the priors (which includes our a priori knowledge about the model parameters) and $\boldsymbol{P(d)}$ is a normalising constant. Bayesian inference allows the model predictions to be projected into observational space, where the measurement uncertainties are defined. This is particularly important when the observational uncertainties are both heteroskedastic and correlated, as is the case here.

We consider three different models for the velocity dispersion of each group, approximating their distributions as Gaussians\footnote{While many of the distributions appear non-Gaussian, a Gaussian distribution still represents a reasonable approximation to the underlying distribution.}. The first is a trivariate Gaussian aligned to the observed coordinate system ($\alpha$, $\delta$, and along the line of sight), hereafter known as the {\it unrotated} model. This model is used because it represents one of the most commonly-used fitting methods employed in the literature\footnote{We note that the results obtained are independent of the, relatively arbitrary, choice of coordinate system, which we verified by performing the same fit in the Galactic Cartesian coordinate system ($UVW$), obtaining consistent results.}. The second model replaces the plane of the sky components with a rotated bivariate Gaussian, alongside the third, line of sight, component (the {\it partially rotated} model). The third model utilises a rotated trivariate Gaussian, allowed to fully rotate in all three dimensions (the {\it fully rotated model}). These models require 6, 7, and 9 model parameters, respectively (3 central velocities, 3 velocity dispersions, and 0, 1, and 3 rotation angles respectively). In addition we utilise an unrotated trivariate Gaussian to represent the non-cluster component (which may include runaway stars or nearby young stars not part of the group under study), requiring a further 6 model parameters for the Gaussians and a seventh parameter, $f_{field}$ to represent the fraction of stars that are not members of the group. For each of the three model cases, a population of $N = 10^5$ stars are modelled with 3D velocities sampled from either the cluster or non-cluster components and projected into the observational space as necessary, assuming all stars are at the same distance (a reasonable assumption given the line-of-sight distance dispersion is small compared to the group distance). For the first two models we use all 2651 stars in our sample for the fits, while for the third model we are limited to the 1867 stars with both RVs and PMs. We repeated the first to the first two models only using the 1867 stars with 3D kinematics and found fully consistent results, albeit with larger uncertainties, showing that this approach does not bias our results.

Unresolved binary systems will broaden the observed RV distribution due to the contribution that binary orbital motion makes to the measured velocity. To simulate this process we follow \citet{oden02} and \citet{cott12} by assuming that a fraction of our sample are in binary systems\footnote{We don't consider triple systems because their properties are poorly constrained and are typically hierarchical, meaning that the third star is usually on a wide, long-period orbit that does not introduce a large velocity offset.}. The fraction of binary stars in young stellar groups is poorly constrained and so we set the binary fraction to be 46\%, appropriate for solar-type field stars \citep{ragh10}. The primary star masses were sampled from a standard mass function \citep{masc13} in the mass range 0.5 -- 1.0 M$_\odot$ (appropriate for the typical stars observed by GES), while the secondary masses were selected from a power-law companion mass ratio distribution with index $\gamma = -0.5$ over the range of mass ratios $q = 0.1$--$1.0$ \citep{regg11}. The orbital periods were selected from a log-normal distribution with a mean period of 5.03 and a dispersion of 2.28 in log$_{10}$ days \citep{ragh10}. The eccentricities were selected from a flat distribution from $e = 0$ to a maximum that scales with the orbital period as proposed by \citet{park09}. We then calculate instantaneous velocity offsets for the primary and secondary stars (relative to the centre of mass of the system) at a random phase within the binary's orbit, and then use the luminosity-weighted average of the two as the photo-centre velocity\footnote{This compensates for the fact that for high mass-ratio binaries some of the light from the secondary contributes to the spectral features used to measure the RV, and thus the measured RV will be intermediate between that of the two stars. From simulations we found that this reduces the broadening of the RV distribution due to binaries by $\sim$25\%.}, which is then added to the modelled RV. Note that we do not correct for the effects of binarity on the measured PMs since this is estimated to be much smaller than the effect on RVs \citep[e.g.,][]{jack20}.

Finally we add measurement uncertainties for the RVs and PMs for each star, randomly sampling these from the observed uncertainty distributions and include the correlated PM uncertainties from {\it Gaia} EDR3. This produces our fully forward-modelled velocity distribution models.

The three models used have 13, 14 and 16 free parameters for the unrotated, partially rotated and fully rotated models, respectively. Wide, uniform, and linear priors were used for each of these parameters covering $0$ to $100$ km~s$^{-1}$ for the velocity dispersions, $-100$ to $+100$ km~s$^{-1}$ for the central velocities, 0 to 0.2 for the field star component and 0 to 180$^\circ$ for the rotation angles.

To sample the posterior distribution function we use the Markov-Chain Monte Carlo (MCMC) ensemble sampler \citep{good13} \textsc{emcee} \citep{fore13} and compare the modelled probability density function to the observations using a 3D unbinned maximum likelihood test, which is made efficient by the smooth velocity distributions modelled. For the MCMC sampler we used 1000 walkers and 2000 iterations, discarding the first half as a burn-in. The parameters were found to have similar autocorrelation lengths, typically of the order of $\sim$200 iterations, resulting in $\sim$10 independent samples per walker. The posterior distribution functions were found to follow a normal distribution, and thus the median value was used as the best fit, with the 16$^{th}$ and 84$^{th}$ percentiles used for the 1$\sigma$ uncertainties.

\begin{table*}
\caption{Velocity dispersion fitting results for the unrotated and partially-rotated models. All PM velocity dispersions have been recalculated in physical units for ease of comparison with the RV dispersions, the uncertainties for which take into account the full distance uncertainties. $\mu_1$ and $\mu_2$ are the velocity dispersions along the semi-major and semi-minor axes of the partially-rotated velocity ellipsoid, the position angle of which is $\theta$.}
\label{table:kinematics} 
\begin{tabular}{lcccclccccc}
\hline
Group		& \multicolumn{4}{c}{Unrotated velocity dispersions} && \multicolumn{5}{c}{Partially rotated velocity dispersions} \\ 
\cline{2-5} \cline{7-11} 
			& RV & $\mu_\alpha$ & $\mu_\delta$ & $\sigma_{3D}$ && RV & $\mu_1$ & $\mu_2$ & $\theta$ & $\sigma_{3D}$ \\
			& [km s$^{-1}$] & [km s$^{-1}$] & [km s$^{-1}$] & [km s$^{-1}$] && [km s$^{-1}$] & [km s$^{-1}$] & [km s$^{-1}$] & [deg.] & [km s$^{-1}$] \\
\hline
Rho Oph & $1.087^{+0.080}_{-0.111}$ & $0.67^{+0.07}_{-0.07}$ & $0.96^{+0.09}_{-0.08}$ & $1.60^{+0.08}_{-0.09}$ && $1.147^{+0.103}_{-0.100}$ & $0.96^{+0.09}_{-0.08}$ & $0.74^{+0.06}_{-0.07}$ & $91^{+24}_{-26}$ &  $1.67^{+0.10}_{-0.08}$\\ 
Cha I (south) & $0.859^{+0.147}_{-0.165}$ & $0.40^{+0.08}_{-0.06}$ & $0.52^{+0.04}_{-0.06}$ & $1.08^{+0.13}_{-0.13}$ && $0.844^{+0.105}_{-0.099}$ & $0.54^{+0.05}_{-0.06}$ & $0.52^{+0.06}_{-0.05}$ & $92^{+30}_{-30}$ &  $1.13^{+0.09}_{-0.08}$\\ 
Cha I (north) & $0.649^{+0.037}_{-0.056}$ & $0.54^{+0.07}_{-0.05}$ & $0.76^{+0.07}_{-0.08}$ & $1.13^{+0.06}_{-0.06}$ && $0.624^{+0.052}_{-0.051}$ & $0.68^{+0.03}_{-0.05}$ & $0.48^{+0.02}_{-0.03}$ & $31^{+6}_{-5}$ &  $1.05^{+0.04}_{-0.05}$\\ 
Gamma Vel & $0.344^{+0.060}_{-0.065}$ & $0.36^{+0.03}_{-0.03}$ & $0.37^{+0.04}_{-0.03}$ & $0.62^{+0.05}_{-0.04}$ && $0.353^{+0.024}_{-0.024}$ & $0.36^{+0.02}_{-0.02}$ & $0.35^{+0.02}_{-0.02}$ & $176^{+6}_{-7}$ &  $0.61^{+0.03}_{-0.02}$\\ 
Vela OB2 & $1.490^{+0.109}_{-0.131}$ & $0.54^{+0.05}_{-0.04}$ & $1.17^{+0.07}_{-0.06}$ & $1.97^{+0.10}_{-0.10}$ && $1.411^{+0.106}_{-0.119}$ & $1.09^{+0.08}_{-0.08}$ & $0.51^{+0.07}_{-0.07}$ & $35^{+6}_{-6}$ &  $1.86^{+0.10}_{-0.10}$\\ 
25 Ori & $0.501^{+0.045}_{-0.048}$ & $0.40^{+0.05}_{-0.04}$ & $0.39^{+0.04}_{-0.04}$ & $0.75^{+0.05}_{-0.04}$ && $0.501^{+0.058}_{-0.061}$ & $0.39^{+0.02}_{-0.02}$ & $0.32^{+0.04}_{-0.03}$ & $193^{+11}_{-11}$ &  $0.71^{+0.05}_{-0.04}$\\ 
Barnard 30 & $0.811^{+0.258}_{-0.206}$ & $1.40^{+0.34}_{-0.22}$ & $1.67^{+0.34}_{-0.26}$ & $2.33^{+0.38}_{-0.20}$ && $0.831^{+0.121}_{-0.150}$ & $1.94^{+0.57}_{-0.52}$ & $0.85^{+0.25}_{-0.19}$ & $47^{+14}_{-12}$ &  $2.27^{+0.53}_{-0.41}$\\ 
$\lambda$ Ori & $0.717^{+0.052}_{-0.048}$ & $0.52^{+0.06}_{-0.04}$ & $0.33^{+0.04}_{-0.03}$ & $0.95^{+0.06}_{-0.04}$ && $0.696^{+0.038}_{-0.036}$ & $0.45^{+0.04}_{-0.04}$ & $0.30^{+0.04}_{-0.03}$ & $136^{+7}_{-7}$ &  $0.88^{+0.04}_{-0.04}$\\ 
Barnard 35 & $1.053^{+0.120}_{-0.105}$ & $0.85^{+0.05}_{-0.06}$ & $0.46^{+0.09}_{-0.06}$ & $1.43^{+0.10}_{-0.08}$ && $1.058^{+0.122}_{-0.109}$ & $0.83^{+0.06}_{-0.05}$ & $0.42^{+0.12}_{-0.04}$ & $97^{+29}_{-5}$ &  $1.41^{+0.12}_{-0.08}$\\ 
S Mon Cluster & $1.993^{+0.146}_{-0.117}$ & $0.59^{+0.06}_{-0.06}$ & $0.58^{+0.05}_{-0.04}$ & $2.16^{+0.14}_{-0.11}$ && $1.903^{+0.083}_{-0.080}$ & $0.88^{+0.24}_{-0.25}$ & $0.64^{+0.12}_{-0.12}$ & $172^{+20}_{-19}$ &  $2.19^{+0.14}_{-0.11}$\\ 
Spokes Cluster & $2.106^{+0.128}_{-0.131}$ & $1.48^{+0.07}_{-0.07}$ & $0.92^{+0.07}_{-0.08}$ & $2.73^{+0.11}_{-0.11}$ && $2.011^{+0.078}_{-0.084}$ & $1.70^{+0.09}_{-0.08}$ & $0.56^{+0.13}_{-0.12}$ & $93^{+1}_{-1}$ &  $2.69^{+0.09}_{-0.08}$\\ 
ASSC 50 & $0.573^{+0.037}_{-0.030}$ & $1.16^{+0.09}_{-0.07}$ & $1.05^{+0.08}_{-0.06}$ & $1.67^{+0.08}_{-0.06}$ && $0.554^{+0.044}_{-0.042}$ & $1.69^{+0.09}_{-0.10}$ & $0.16^{+0.03}_{-0.03}$ & $56^{+3}_{-3}$ &  $1.79^{+0.09}_{-0.09}$\\ 
Col. 197 & $0.655^{+0.035}_{-0.040}$ & $1.15^{+0.10}_{-0.09}$ & $1.19^{+0.12}_{-0.09}$ & $1.78^{+0.11}_{-0.08}$ && $0.657^{+0.032}_{-0.030}$ & $1.12^{+0.11}_{-0.12}$ & $0.77^{+0.10}_{-0.07}$ & $47^{+7}_{-8}$ &  $1.51^{+0.11}_{-0.09}$\\ 
NGC 6530 & $2.242^{+0.218}_{-0.222}$ & $2.51^{+0.14}_{-0.13}$ & $1.80^{+0.09}_{-0.09}$ & $3.82^{+0.17}_{-0.15}$ && $2.320^{+0.251}_{-0.264}$ & $2.63^{+0.12}_{-0.13}$ & $2.63^{+0.14}_{-0.13}$ & $100^{+20}_{-19}$ &  $4.38^{+0.18}_{-0.17}$\\ 
NGC 2244 & $2.658^{+0.018}_{-0.015}$ & $1.69^{+0.27}_{-0.36}$ & $1.78^{+0.23}_{-0.28}$ & $3.62^{+0.18}_{-0.20}$ && $2.698^{+0.063}_{-0.062}$ & $1.26^{+0.13}_{-0.13}$ & $1.23^{+0.13}_{-0.14}$ & $102^{+11}_{-11}$ &  $3.22^{+0.09}_{-0.09}$\\ 
NGC 2237 & $0.492^{+0.071}_{-0.063}$ & $1.00^{+0.41}_{-0.21}$ & $1.10^{+0.37}_{-0.26}$ & $1.56^{+0.43}_{-0.20}$ && $0.501^{+0.068}_{-0.068}$ & $1.14^{+0.20}_{-0.27}$ & $1.12^{+0.21}_{-0.26}$ & $46^{+32}_{-32}$ &  $1.67^{+0.21}_{-0.23}$\\ 
Trumpler 14 & $6.084^{+0.234}_{-0.192}$ & $3.07^{+0.15}_{-0.16}$ & $3.13^{+0.19}_{-0.18}$ & $7.50^{+0.22}_{-0.18}$ && $6.025^{+0.193}_{-0.161}$ & $3.12^{+0.18}_{-0.16}$ & $2.85^{+0.13}_{-0.13}$ & $81^{+8}_{-8}$ &  $7.36^{+0.19}_{-0.15}$\\ 
Trumpler 16 & $2.879^{+0.146}_{-0.139}$ & $3.26^{+0.30}_{-0.27}$ & $2.29^{+0.21}_{-0.19}$ & $4.91^{+0.25}_{-0.20}$ && $2.724^{+0.181}_{-0.166}$ & $2.97^{+0.23}_{-0.28}$ & $1.80^{+0.13}_{-0.14}$ & $163^{+5}_{-5}$ &  $4.41^{+0.20}_{-0.21}$\\ 

\\ \hline
\end{tabular}
\end{table*}

\subsection{Velocity dispersion fitting results}
\label{s-velocityfits}

Velocity dispersion fitting was performed for all 18 groups using all three models described above. Table \ref{table:kinematics} lists the fitted 1D velocity dispersions determined using the unrotated and partially-rotated velocity dispersion models. The 1D velocity dispersions fitted using these methods vary from 0.2 -- 6.1 km~s$^{-1}$, with the majority around $\sim$1 km~s$^{-1}$. Example forward-modelled fits are shown in Figure~\ref{velocity_dispersion_fit_nonrotated} for the unrotated model for the Gamma Velorum cluster and in Figure~\ref{velocity_dispersion_fit_2Drotated} for the partially rotated model for ASSC~50.

For the fully-rotated model, not all fits provided an improvement on the partially-rotated model fits. Bayesian Information Criterion \citep[BIC,][]{schw78} was used to assess the improvement in the model fit achieved for the more complex models (those with partial or full rotation of the velocity ellipsoid) by applying a penalty to the likelihood of the more complex models compared to the unrotated model. If the BIC indicated that the more complex model did not provide a sufficiently better fit to the data than the simple model, then the former models were not considered. This typically happened when the model fitting process could not identify a rotation angle at which the velocity distributions were better fit than with an unrotated model.

\begin{figure*}
\centering
\includegraphics[width=440pt,trim=0 0 0 0]{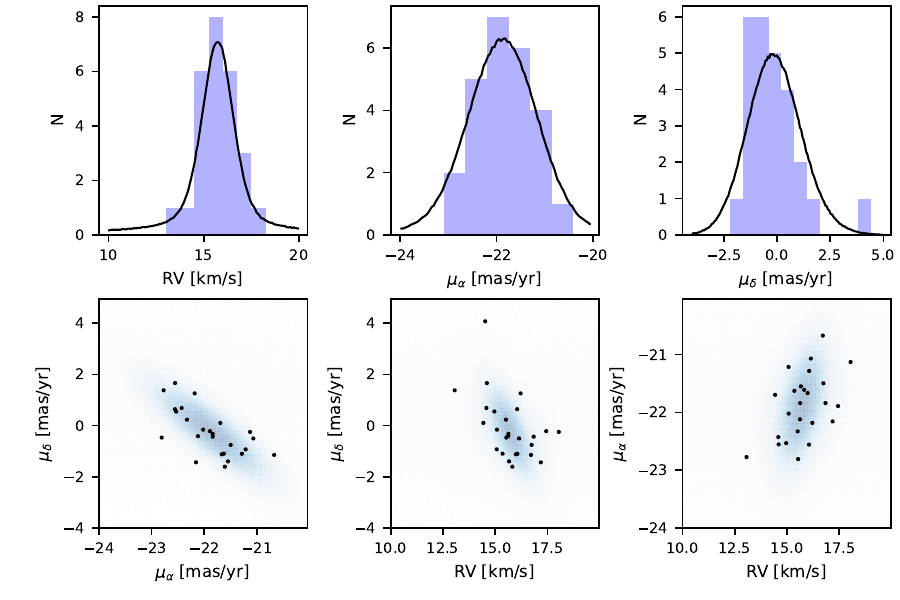}
\caption{Velocity distributions for young stars in Cha~I North (45 stars with RVs, 26 stars with PMs) with our forward-modelled, fully-rotated velocity dispersion fits superimposed. The top panels show the three velocity distributions (blue histograms) with projected velocity dispersion fits (black lines), while the bottom panels show the three 2D projections of the velocity distributions (black points) with the projected velocity dispersion fits as a blue 2D histogram.}
\label{velocity_dispersion_fit_3Drotated}
\end{figure*}

Table~\ref{table:3D} lists the five groups for which the fully-rotated model provided significantly improved fits to the data compared to the partially-rotated model. The groups for which this was the case are all nearby (within 500~pc) and therefore low-mass. This leads to a bias wherein the 1D velocity dispersions are lower than for the full sample, typically less than 1~km~s$^{-1}$. Figure~\ref{velocity_dispersion_fit_3Drotated} shows the fully-rotated model fit for the cluster Cha~I~North.

\begin{table*}
\caption{Velocity dispersion fit results for the fully-rotated model where the fit provides an improvement on the partially-rotated model fit (Table~\ref{table:kinematics}), assessed using BIC. $v_1$, $v_2$ and $v_2$ are the velocity dispersions along the three axes, with rotation angles of $\theta$, $\psi$ and $\phi$.}
\label{table:3D} 
\begin{tabular}{lccccccc}
\hline
Group			& \multicolumn{7}{c}{Fully rotated velocity dispersions} \\
\cline{2-8}
				& $v_1$ & $v_2$ & $v_3$ & $\theta$	 & $\psi$		 & $\phi$	 & $\sigma_{3D}$ \\
				& [km s$^{-1}$] & [km s$^{-1}$] & [km s$^{-1}$] & [deg.] & [deg.] & [deg.] & [km s$^{-1}$] \\
\hline
Rho Oph & $0.85^{+0.08}_{-0.07}$ & $0.35^{+0.07}_{-0.05}$ & $0.91^{+0.09}_{-0.07}$&$39^{+14}_{-18}$ & $40^{+15}_{-15}$ & $3^{+1}_{-1}$ & $1.30^{+0.09}_{-0.06}$ \\ 
Cha I (south) & $1.01^{+0.24}_{-0.17}$ & $0.39^{+0.04}_{-0.02}$ & $0.06^{+0.02}_{-0.01}$&$33^{+19}_{-15}$ & $54^{+22}_{-19}$ & $53^{+18}_{-16}$ & $1.09^{+0.23}_{-0.15}$ \\ 
Cha I (north) & $1.32^{+0.11}_{-0.18}$ & $0.25^{+0.09}_{-0.04}$ & $0.42^{+0.08}_{-0.06}$&$45^{+20}_{-15}$ & $54^{+14}_{-10}$ & $83^{+18}_{-20}$ & $1.41^{+0.12}_{-0.16}$ \\ 
Vela OB2 & $1.40^{+0.10}_{-0.10}$ & $0.60^{+0.07}_{-0.07}$ & $0.80^{+0.11}_{-0.13}$&$38^{+8}_{-8}$ & $40^{+6}_{-6}$ & $43^{+10}_{-10}$ & $1.72^{+0.10}_{-0.10}$ \\ 
Barnard 30 & $0.45^{+0.06}_{-0.05}$ & $0.66^{+0.06}_{-0.06}$ & $1.54^{+0.20}_{-0.28}$&$19^{+29}_{-13}$ & $24^{+31}_{-17}$ & $46^{+11}_{-10}$ & $1.73^{+0.18}_{-0.24}$ \\ 

\\ \hline
\end{tabular}
\end{table*}

The 3D velocity dispersions calculated using the three different models generally agree very well with each other. 61\% of the unrotated model fits agree with the partially-rotated model fits within 1$\sigma$ and 94\% within 2$\sigma$, approximately as expected for a normal distribution. The agreement is poorer between the fully-rotated model fits and the partially-rotated model fits, with only 40\% within 1$\sigma$ and 60\% within 2$\sigma$. The more complex models do give consistently smaller 3D velocity dispersions than the less complex models. Notably, the fully-rotated model fitted 3D velocity dispersions are, on average, 10\% smaller than the unrotated 3D velocity dispersions. This implies that velocity dispersions previously calculated without considering the rotation of the velocity ellipsoid may have over-estimated both the 1D and 3D velocity dispersions by $\sim$10\%.

Regardless of how the velocity dispersions are measured they are significantly anisotropic. The fraction of groups whose velocity dispersions are inconsistent with isotropy to 1$\sigma$ is 94\% for the unrotated models, 89\% for the partially-rotated models, and 100\% for the fully-rotated models. Limiting this comparison to just the PM axis velocity dispersions (to remove any influence of binarity or distance uncertainty) then the fractions are slightly lower at 50\% for the unrotated models and 67\% for the partially-rotated models. Despite this it is clear that the vast majority of these groups do not have isotropic velocity dispersions.

Our results are generally in good agreement with previous studies when comparing like-for-like, albeit with a slight tendency to recover lower velocity dispersions. Comparing our results to the 1D RV dispersions calculated by \citet{jeff14} and \citet{rigl16} for the Gamma Vel cluster, Vela OB2 and Rho Ophiuchus, we find excellent agreement (i.e., within 1 $\sigma$), as would be expected given that the RV data is the same (though it has been updated as the GES pipeline has improved). Our velocity dispersions for Cha~I North and South are also consistent with the velocity dispersions of $0.95 \pm 0.18$ and $0.87 \pm 0.24$ km~s$^{-1}$ calculated by \citet{sacc17}, respectively. Our results are in reasonable agreement with the 2D PM velocity dispersions calculated by \citet{kuhn19} for the S~Mon cluster, NGC~2244, NGC~6530, Trumpler 14 and Trumpler 16, most agreeing within 1--2 $\sigma$, though our velocity dispersions are, on average, slightly smaller. This could be due to differences in the group samples or membership \citep[][did not have access to spectroscopic data to confirm the youth of their sample]{kuhn19}. Finally, our 3D velocity distribution is smaller than that calculated by \citet{wrig16} for NGC~6530, despite using similar data. This can be attributed to different membership criteria since \citet{wrig19} used a combination of spectroscopic and X-ray youth indicators, while we have only used spectroscopy.

\subsection{Virial state}
\label{s-virialstate}

To measure the virial state of each group we use the virial equation, which in its three-dimensional form is given by

\begin{equation}
\sigma_{3D, vir}^2 = \frac{G \, M}{2 \, r_{vir}}
\label{eqn-virial}
\end{equation}

\noindent where $\sigma_{3D}$ is the 3D velocity dispersion, $G$ is the gravitational constant, $M$ is the mass and $r_{vir}$ is the radius. We substitute the parameter $\eta = 6 r_{vir} / r_{eff}$, where $r_{eff}$ is the effective (or half-light) radius in projection, to get

\begin{equation}
\sigma_{3D, vir}^2 = \frac{3 \, G \, M_{vir}}{\eta \, r_{eff}}
\label{equation:virial}
\end{equation}

\noindent The parameter $\eta$ can be derived from the power-law index, $\gamma$, of an \citet[][hereafter EFF]{elso87} surface brightness profile \citep[see e.g.,][]{port10}, from which the effective radius, $r_{eff}$, can also be measured. Some studies assume a standard value of $\eta = 10$, a reasonable approximation for young clusters, but one that hides a significant level of uncertainty in the true value for a given group or cluster. We used the ancillary datasets mentioned in Section~2 to fit EFF profiles for all our groups, deriving $r_{eff}$ and $\gamma$, with uncertainties calculated using a bootstrapping process. The power-law indexes, $\gamma$, and therefore also the $\eta$ values, are not well constrained for these groups, with values ranging from $\eta = 6$--11, with varying levels of precision (see Table~\ref{table:virial}). There are no direct measurements of $\eta$ or $\gamma$ in the literature for these clusters (or almost any cluster for that matter), so this is difficult to check and represents one of the main uncertainties in virial calculations.

The fitted values of $r_{eff}$ are also listed in Table~\ref{table:virial}. These are generally in good agreement with measurements from the literature, though different studies often define or measure the radius in different ways. For example, for Cha~I North and South, \citet{luhm07} estimate radii of $\sim$0.1 and $\sim$0.2 deg. ($\sim$0.3 and $\sim$0.6 pc) respectively, while we measure radii of 0.42 and 0.40 pc, in broad agreement. For the Spokes Cluster, \citet{maiz19} estimate a full diameter of $\sim$1.3~pc, which is consistent with our effective radius of $0.39^{+0.03}_{-0.03}$~pc, while for Trumpler~14 \citet{fige08} measure an effective radius of $\sim$0.5~pc, consistent with our effective radius of $0.53^{+0.05}_{-0.07}$~pc.

The effective radii are combined with the stellar and gas masses listed in Section~\ref{s_targets} and Equation \ref{equation:virial} to give virial velocity dispersions. Table~\ref{table:virial} compares these values with our measured velocity dispersions from Tables \ref{table:kinematics} and \ref{table:3D}. For ease of comparison we also provide in Table~\ref{table:virial} values of the virial ratio, $\alpha = \sigma_{3D} / \sigma_{3D, vir}$, for which $\alpha < 1$ indicates a sub-virial system, $\alpha = 1$ indicates a system in virial equilibrium, and $\alpha > 1$ indicates a super-virial system. Where relevant we have calculated virial velocity dispersions and ratios using just the stellar mass and the sum of stellar and gas masses.

\begin{table*}
\caption{Velocity dispersions, stellar and gas masses, effective radii and virial states for all groups. The 3D velocity dispersion, $\sigma_{3D}$, is the partially-rotated model from Table~\ref{table:kinematics} unless an improved fit was achieved using the fully-rotated model from Table~\ref{table:3D}. The stellar and molecular gas masses were gathered from the literature, as described in Section~\ref{s_targets}. The effective radii were fitted as described in the text. The virial velocity dispersions were calculated using Equation~\ref{equation:virial} using either the stellar mass or the sum of the stellar and gas masses. The virial ratio is given by $\alpha = \sigma_{3D} / \sigma_{3D, vir}$, again using either the stellar mass or the sum of the stellar and gas masses.}
\label{table:virial} 
\begin{tabular}{lcccccccccc}
\hline
Group		& $\sigma_{3D}$	& \multicolumn{2}{c}{$M$ [$M_\odot$]} & $r_{eff}$	& $\eta$	& \multicolumn{2}{c}{$\sigma_{3D, vir}$ [km s$^{-1}$]}	&& \multicolumn{2}{c}{$\alpha$} \\
\cline{3-4} \cline{7-8} \cline{10-11}
			& [km s$^{-1}$]		& (stars)			& (gas)		& [pc]		&		& (stars) & (stars $+$ gas) && (stars) & (stars $+$ gas) \\
\hline
Rho Oph & $1.30^{+0.09}_{-0.06}$ & 106 & 1750 & 0.69$^{+0.08}_{-0.08}$ & 10.1$^{+0.3}_{-0.3}$ & 0.44$^{+0.01}_{-0.04}$ & 1.86$^{+0.06}_{-0.16}$ && 2.93$^{+0.38}_{-0.13}$ & 0.70$^{+0.09}_{-0.03}$ \\ 
Cha I (south) & $1.09^{+0.23}_{-0.15}$ & 59 & 1000 & 0.4$^{+0.1}_{-0.09}$ & 6.0$^{+0.1}_{-3.5}$ & 0.56$^{+0.05}_{-0.16}$ & 2.39$^{+0.19}_{-0.68}$ && 1.93$^{+1.00}_{-0.24}$ & 0.46$^{+0.24}_{-0.06}$ \\ 
Cha I (north) & $1.41^{+0.12}_{-0.16}$ & 54 & 1000 & 0.42$^{+0.15}_{-0.19}$ & 6.0$^{+0.5}_{-3.5}$ & 0.53$^{+0.01}_{-0.20}$ & 2.33$^{+0.01}_{-0.88}$ && 2.67$^{+1.69}_{-0.01}$ & 0.60$^{+0.38}_{-0.01}$ \\ 
Gamma Vel & $0.61^{+0.03}_{-0.02}$ & 152 & - & 1.88$^{+0.27}_{-0.23}$ & 10.6$^{+1.0}_{-3.6}$ & 0.31$^{+0.10}_{-0.01}$ & - && 1.96$^{+0.08}_{-0.47}$ & - \\ 
Vela OB2 & $1.72^{+0.10}_{-0.10}$ & 1285 & - & 23.5$^{+1.1}_{-2.0}$ & 9.5$^{+0.1}_{-0.1}$ & 0.27$^{+0.02}_{-0.01}$ & - && 6.32$^{+0.25}_{-0.58}$ & - \\ 
25 Ori & $0.71^{+0.05}_{-0.04}$ & 400 & - & 2.55$^{+0.3}_{-0.18}$ & 9.8$^{+0.5}_{-1.9}$ & 0.46$^{+0.02}_{-0.04}$ & - && 1.56$^{+0.18}_{-0.11}$ & - \\ 
Barnard 30 & $1.73^{+0.18}_{-0.24}$ & 60 & - & 1.75$^{+0.8}_{-1.71}$ & 8.7$^{+0.1}_{-3.5}$ & 0.23$^{+0.01}_{-0.09}$ & - && 7.68$^{+5.74}_{-0.10}$ & - \\ 
$\lambda$ Ori & $0.88^{+0.04}_{-0.04}$ & 650 & - & 0.23$^{+0.29}_{-0.96}$ & 8.7$^{+1.1}_{-3.9}$ & 2.06$^{+0.01}_{-1.39}$ & - && 0.43$^{+0.92}_{-0.01}$ & - \\ 
Barnard 35 & $1.41^{+0.12}_{-0.08}$ & 73 & - & 4.66$^{+0.8}_{-0.93}$ & 9.5$^{+0.1}_{-3.5}$ & 0.15$^{+0.07}_{-0.01}$ & - && 9.64$^{+0.42}_{-3.04}$ & - \\ 
S Mon Cluster & $2.19^{+0.14}_{-0.11}$ & 425 & 3000 & 0.17$^{+0.05}_{-0.09}$ & 6.0$^{+1.2}_{-3.5}$ & 2.33$^{+0.01}_{-0.96}$ & 6.62$^{+0.01}_{-2.74}$ && 0.94$^{+0.68}_{-0.01}$ & 0.33$^{+0.24}_{-0.01}$ \\ 
Spokes Cluster & $2.69^{+0.09}_{-0.08}$ & 425 & 3000 & 0.39$^{+0.03}_{-0.03}$ & 10.6$^{+2.0}_{-2.7}$ & 1.15$^{+0.45}_{-0.01}$ & 3.27$^{+1.26}_{-0.03}$ && 2.34$^{+0.03}_{-0.66}$ & 0.82$^{+0.01}_{-0.23}$ \\ 
ASSC 50 & $1.79^{+0.09}_{-0.09}$ & 200 & - & 4.24$^{+0.37}_{-0.28}$ & 6.0$^{+3.6}_{-0.1}$ & 0.32$^{+0.01}_{-0.06}$ & - && 5.62$^{+1.35}_{-0.26}$ & - \\ 
Col. 197 & $1.51^{+0.11}_{-0.09}$ & 79 & - & 2.64$^{+0.55}_{-0.44}$ & 9.6$^{+0.1}_{-0.1}$ & 0.20$^{+0.01}_{-0.02}$ & - && 7.52$^{+1.23}_{-0.44}$ & - \\ 
NGC 6530 & $4.38^{+0.18}_{-0.17}$ & 3125 & 40000 & 1.49$^{+0.17}_{-0.19}$ & 6.0$^{+3.5}_{-0.1}$ & 2.13$^{+0.06}_{-0.58}$ & 7.90$^{+0.21}_{-2.15}$ && 2.06$^{+0.78}_{-0.07}$ & 0.55$^{+0.21}_{-0.02}$ \\ 
NGC 2244 & $3.22^{+0.09}_{-0.09}$ & 1300 & - & 1.19$^{+0.1}_{-0.24}$ & 6.0$^{+4.7}_{-0.1}$ & 1.54$^{+0.07}_{-0.40}$ & - && 2.10$^{+0.75}_{-0.11}$ & - \\ 
NGC 2237 & $1.67^{+0.21}_{-0.23}$ & 250 & - & 2.07$^{+0.54}_{-0.52}$ & 6.0$^{+1.2}_{-3.5}$ & 0.51$^{+0.09}_{-0.14}$ & - && 3.29$^{+1.42}_{-0.65}$ & - \\ 
Trumpler 14 & $7.36^{+0.19}_{-0.15}$ & 5400 & - & 0.53$^{+0.05}_{-0.07}$ & 6.0$^{+0.3}_{-4.7}$ & 4.70$^{+0.19}_{-1.30}$ & - && 1.57$^{+0.61}_{-0.06}$ & - \\ 
Trumpler 16 & $4.41^{+0.20}_{-0.21}$ & 3250 & - & 1.72$^{+0.12}_{-0.11}$ & 9.5$^{+0.2}_{-0.1}$ & 1.60$^{+0.08}_{-0.06}$ & - && 2.75$^{+0.17}_{-0.18}$ & - \\ 

\\ \hline
\end{tabular}
\end{table*}

When considering only the contribution of the stellar mass to the gravitational potential, nearly all of the groups are super-virial ($\alpha > 1$), with only $\lambda$~Ori and the S Mon cluster in NGC~2264 being in virial equilibrium or sub-virial ($\alpha < 1$). A small number of systems are close to being in virial equilibrium with $\alpha < 2$, such as Cha~I South, Gamma~Vel, 25~Ori, and Trumpler 14. For these clusters if their stellar mass has been under-estimated or their radii or $\eta$ values over-estimated, then this may bring them into virial equilibrium\footnote{Note that these calculations do not take into account the uncertainty on the stellar mass, which is difficult to quantify and could be as high as 20--30\% for some of these systems.}. We note that 25~Ori and the Gamma Vel cluster are highly clustered and sufficiently old (19 and 19.5 Myr, respectively) that we would expect them to be gravitationally bound (if they were not then they would have expanded and dispersed), which suggests that certainly these clusters have their masses under-estimated or $r_{eff}$ or $\eta$ over-estimated.

A different picture emerges if we consider the gravitational potential resulting from both the stellar and gas parts of the local system. Molecular gas masses were gathered from the literature for any group still associated with or embedded within a molecular cloud and for which such data were available (as described in Section~\ref{s_targets}). This was the case for six groups, Rho Oph, Cha~I North and South, the two clusters in NGC~2264 and NGC~6530. We note that certain other groups are associated with molecular gas, but estimates of the total gas mass were not available from the literature (Barnard 30 and 35, ASSC 50, NGC 2244, Trumpler 14 and 16), while the remaining groups are not associated with any molecular gas (Gamma Vel, Vela OB2, 25 Ori, $\lambda$ Ori, Collinder 197 and NGC 2237). When the molecular gas mass is taken into account, all of the groups with such information available are found to be sub-virial, with typical virial ratios of $\sim$0.5. Some of these groups may not be fully embedded within their molecular cloud, or the molecular gas may be more spatially extended than the stellar part of the system, both of which would mean that these gas masses would be over-estimates in this context. Nonetheless, it is notable that when the molecular gas masses in these regions are taken into account, all such systems are found to be sub-virial and some would still be in virial equilibrium if the gravitational potentials were half that estimated here.

\section{Group expansion}

Many young stellar systems, particularly OB associations, have recently been found to be expanding \citep{cant19b,arms20,wrig20}. Younger star forming regions have presented mixed results, with some young groups expanding, while others are not \citep[e.g.,][]{kuhn19}. Inspection of velocity vector maps of these groups show many tend to exhibit a preference for coherent outward motion, particularly NGC 6530, Vela OB2, and ASCC~50. In this section we quantify the level of expansion in the groups in our sample using a variety of measures, correcting all transverse velocities for radial streaming motions (virtual expansion) using their central radial velocities and equation A3 in \citet{brow97}, before performing this analysis.

\begin{table*}
\caption{Expansion gradients and indicators of expansion for the groups studied in this work. 1D expansion gradients were calculated in RA ($\alpha$), Dec ($\delta$) and along the line of sight ($\varpi$). Rotated expansion gradients were calculated in the plane of the sky and rotated in steps of 5$^\circ$ until the largest single expansion gradient (positive or negative) was found. The gradient in this direction and the gradient perpendicular to this direction are given, as well as the angle of maximum expansion gradient. Weighted-median expansion velocities are calculated from the radial component of the 2D (plane of the sky) velocities of all stars in each group. All transverse velocities were corrected for the effects of radial streaming motions.}
\label{expansion_table} 
\begin{tabular}{l ccc c ccc c c}
\hline
Group & \multicolumn{3}{c}{1D expansion gradients} && \multicolumn{3}{c}{Rotated expansion gradients} && $\widetilde{v}_{out}$ \\
\cline{2-4} \cline{6-8}
	     & $\varpi$ & $\alpha$ & $\delta$ && $x_1$ & $x_2$ & $\theta$ \\
	     & [km/s/pc] & [km/s/pc] & [km/s/pc] && [km/s/pc] & [km/s/pc] & [$^\circ$] && [km/s]\\
\hline
Rho Oph & $-0.061^{+0.138}_{-0.309}$ & $-0.14^{+0.21}_{-0.20}$ & $-0.54^{+0.24}_{-0.22}$ && $-0.18^{+0.22}_{-0.20}$ & $-0.53^{+0.24}_{-0.21}$ & 50 && $-0.34^{+0.01}_{-0.12}$ \\ [5pt] 
Cha I (south) & $-0.016^{+0.054}_{-0.234}$ & $0.28^{+0.20}_{-0.26}$ & $0.07^{+0.14}_{-0.12}$ && $0.46^{+0.18}_{-0.20}$ & $0.19^{+0.18}_{-0.18}$ & 170 && $0.10^{+0.13}_{-0.08}$ \\ [5pt] 
Cha I (north) & $0.21^{+0.17}_{-0.19}$ & $0.28^{+0.12}_{-0.14}$ & $0.40^{+0.24}_{-0.15}$ && $0.56^{+0.12}_{-0.15}$ & $0.29^{+0.11}_{-0.11}$ & 90 && $0.29^{+0.16}_{-0.10}$ \\ [5pt] 
Gamma Vel &  - & $0.00^{+0.04}_{-0.04}$ & $0.01^{+0.04}_{-0.04}$ && $0.05^{+0.04}_{-0.04}$ & $0.01^{+0.03}_{-0.03}$ & 15 && $-0.06^{+0.06}_{-0.01}$ \\ [5pt] 
Vela OB2 & $0.046^{+0.020}_{-0.048}$ & $-0.14^{+0.09}_{-0.09}$ & $0.43^{+0.10}_{-0.12}$ && $0.43^{+0.11}_{-0.12}$ & $-0.03^{+0.06}_{-0.07}$ & 75 && $0.48^{+0.01}_{-0.08}$ \\ [5pt] 
25 Ori & $0.006^{+0.006}_{-0.008}$ & $0.11^{+0.03}_{-0.03}$ & $0.01^{+0.01}_{-0.01}$ && $0.12^{+0.03}_{-0.03}$ & $0.07^{+0.04}_{-0.04}$ & 50 && $0.16^{+0.03}_{-0.01}$ \\ [5pt] 
Barnard 30 &  - & $0.38^{+0.13}_{-0.14}$ & $-0.13^{+0.10}_{-0.13}$ && $0.39^{+0.11}_{-0.12}$ & $-0.12^{+0.09}_{-0.09}$ & 0 && $0.13^{+0.16}_{-0.11}$ \\ [5pt] 
$\lambda$ Ori &  - & $0.05^{+0.04}_{-0.04}$ & $0.15^{+0.03}_{-0.04}$ && $0.20^{+0.03}_{-0.04}$ & $-0.01^{+0.03}_{-0.02}$ & 75 && $0.24^{+0.06}_{-0.01}$ \\ [5pt] 
Barnard 35 &  - & $0.50^{+0.11}_{-0.14}$ & $0.43^{+0.10}_{-0.12}$ && $0.79^{+0.08}_{-0.14}$ & $0.57^{+0.11}_{-0.13}$ & 120 && $0.61^{+0.02}_{-0.05}$ \\ [5pt] 
S Mon Cluster &  - & $0.07^{+0.04}_{-0.04}$ & $0.20^{+0.03}_{-0.04}$ && $0.35^{+0.04}_{-0.04}$ & $0.11^{+0.03}_{-0.04}$ & 120 && $0.40^{+0.01}_{-0.06}$ \\ [5pt] 
Spokes Cluster &  - & $-0.02^{+0.06}_{-0.06}$ & $0.19^{+0.04}_{-0.05}$ && $0.24^{+0.04}_{-0.05}$ & $-0.02^{+0.06}_{-0.06}$ & 85 && $0.23^{+0.03}_{-0.07}$ \\ [5pt] 
ASSC 50 &  - & $0.21^{+0.03}_{-0.03}$ & $0.23^{+0.02}_{-0.03}$ && $0.27^{+0.02}_{-0.02}$ & $0.15^{+0.03}_{-0.03}$ & 65 && $1.16^{+0.08}_{-0.06}$ \\ [5pt] 
Col. 197 &  - & $0.27^{+0.10}_{-0.12}$ & $0.60^{+0.12}_{-0.13}$ && $0.59^{+0.12}_{-0.13}$ & $0.29^{+0.10}_{-0.11}$ & 85 && $0.59^{+0.08}_{-0.12}$ \\ [5pt] 
NGC 6530 &  - & $-0.05^{+0.06}_{-0.06}$ & $0.40^{+0.06}_{-0.07}$ && $0.49^{+0.06}_{-0.06}$ & $-0.03^{+0.04}_{-0.04}$ & 80 && $0.76^{+0.10}_{-0.02}$ \\ [5pt] 
NGC 2244 &  - & $0.37^{+0.11}_{-0.13}$ & $-0.01^{+0.06}_{-0.07}$ && $0.62^{+0.14}_{-0.14}$ & $0.16^{+0.11}_{-0.12}$ & 140 && $0.88^{+0.07}_{-0.23}$ \\ [5pt] 
NGC 2237 &  - & $0.05^{+0.18}_{-0.18}$ & $0.90^{+0.45}_{-0.33}$ && $0.93^{+0.46}_{-0.35}$ & $0.05^{+0.17}_{-0.17}$ & 90 && $0.16^{+0.39}_{-0.13}$ \\ [5pt] 
Trumpler 14 &  - & $-0.13^{+0.18}_{-0.17}$ & $-0.99^{+0.33}_{-0.29}$ && $0.12^{+0.10}_{-0.10}$ & $-0.13^{+0.21}_{-0.21}$ & 85 && $0.07^{+0.01}_{-0.45}$ \\ [5pt] 
Trumpler 16 &  - & $0.20^{+0.19}_{-0.19}$ & $0.27^{+0.14}_{-0.15}$ && $0.49^{+0.19}_{-0.21}$ & $0.23^{+0.13}_{-0.14}$ & 30 && $1.37^{+0.09}_{-0.28}$ \\ [5pt] 

\\ \hline
\end{tabular}
\end{table*}

\subsection{Fitting 1D expansion gradients}
\label{s-1D_expansion}

The traditional method of measuring the expansion of a group of stars is to search for velocity gradients, i.e., correlations between position and velocity in the same axis that would suggest a ballistic expansion. We follow the method of \citet{wrig18} and fit linear relationships of the form $v = A x + B$ between the velocity, $v$, and spatial position, $x$, in each dimension ($\alpha$, $\delta$, and along the line of sight, $\varpi$, if the group is resolved). The gradient, $A$, and zero-point, $B$, were fitted by maximising the likelihood function, using the MCMC ensemble sampler \textsc{emcee} to sample the posterior distribution. A third parameter ($f$) was introduced to represent the scatter in the relationship \citep[see][]{hogg10} and which was marginalised over to calculate the fit and uncertainties on the fitted velocity gradients.

The results of the velocity gradient fits are listed in Table~\ref{expansion_table}. The fitted expansion gradients vary from $-0.99^{+0.33}_{-0.29}$ (indicating contraction) up to $0.90^{+0.45}_{-0.33}$ km~s$^{-1}$~pc$^{-1}$ (indicating expansion), for Trumpler 14 and NGC~2237 respectively. These values are broadly consistent with, but typically larger than, estimates of expansion made for other groups such as OB associations, which typically extend up to 0.1--0.2 km~s$^{-1}$~pc$^{-1}$ \citep{wrig18,quin21,arms22}. We explored the effects of removing 2 or 3$\sigma$ outliers in position or velocity from the samples before performing the fits, but found that it did not have a significant effect on the results.

\begin{figure}
\centering
\includegraphics[width=240pt,trim=0 0 0 0]{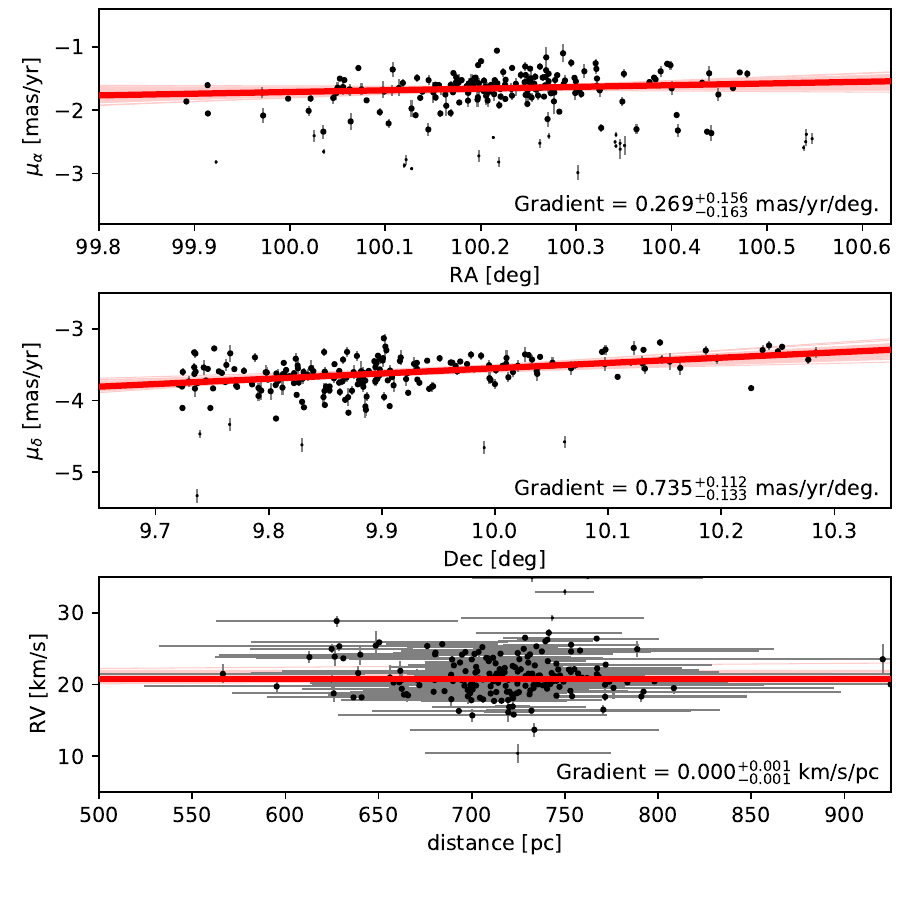}
\caption{Expansion gradients in the S~Mon cluster of NGC~2264 for 200 stars with PMs and 196 stars with RVs. From top to bottom are, respectively, PM in RA plotted against RA, PM in declination plotted against Declination, and RV plotted against the distance. The red line shows the best-fitting 1D expansion gradient, with 100 additional fits sampled from the posterior distribution shown in light red. The best-fitting gradient, with uncertainties are listed in each panel.}
\label{example_expansion_gradient_1D}
\end{figure}

Figure~\ref{example_expansion_gradient_1D} shows an example of a fitted 1D expansion gradients for the S~Mon cluster of NGC~2264, with fitted gradients of $0.27^{+0.16}_{-0.16}$ mas yr$^{-1}$ deg$^{-1}$ = $0.07^{+0.04}_{-0.04}$ km~s$^{-1}$~pc$^{-1}$ in RA and $0.74^{+0.11}_{-0.13}$ mas yr$^{-1}$ deg$^{-1}$ = $0.20^{+0.03}_{-0.04}$ km~s$^{-1}$~pc$^{-1}$ in Dec. These $\sim$1.5 and $\sim$5.5 $\sigma$ measurements of expansion are in strong contrast to the lack of expansion measured along the line of sight (due to the S~Mon cluster not being fully resolved with the available {\it Gaia} data).

Along the line of sight the results are limited. We do not fit or report expansion gradients for the most distant or compact groups studied, where the precision of the parallax measurements do not allow us to probe the line of sight distribution of sources. Of the nearest 6 groups, Gamma Vel is not resolved, and only 3 of the other 5 groups show evidence of expansion, all at only $\sim$1$\sigma$ significance (Cha I North, Vela OB2 and 25~Ori).

In the RA and Dec directions we find more significant evidence for expansion, with 11 out of 18 groups showing evidence for expansion in RA and 11 out of 18 groups showing evidence for expansion in Dec. Notable examples of high-significance indications of expansion are B35 (3$\sigma$ evidence for expansion in both RA and Dec), NGC 6530 \citep[6$\sigma$ evidence for expansion in Dec, but no evidence for expansion in RA, see also][]{wrig19}, the S Mon cluster in NGC 2264 (6$\sigma$ evidence for expansion in Dec, but only 1$\sigma$ in RA), and ASSC 50 (9$\sigma$ evidence for expansion in both RA and Dec). Notably, we also find that Trumpler 14 is contracting at 3$\sigma$ significance in Dec and Rho Oph is contracting at 2$\sigma$ significance in Dec. We will return to these cases later.

In the groups with evidence for expansion in one or both of the plane of the sky directions, approximately half (9 out of 15) have expansion rates that differ significantly ($>$2$\sigma$) between axes (i.e., they have anisotropic expansion), while the remaining 6 are consistent with having isotropic expansion or exhibit anisotropy at only the 1$\sigma$ level. This is in slight contrast with the strong evidence for anisotropic expansion found in OB associations \citep{wrig20}.

\subsection{Fitting rotated expansion gradients}
\label{s_rotated_expansion}

Given that more than half of the groups studied exhibit evidence for anisotropic expansion, and that there is no reason to assume that the strongest expansion would occur along one of our arbitrarily-defined observational axes, we also explored the evidence for expansion along arbitrary axes in the plane of the sky. To achieve this we rotated the 2D plane of the sky PM axes in steps of 5$^\circ$, reprojecting the PMs, and repeating the expansion gradient fits as described in Section~\ref{s-1D_expansion}.

Table~\ref{expansion_table} reports the angle at which the strongest evidence for expansion was found, the expansion gradient fit along that axis, as well as along the perpendicular axis. We fit expansion gradients that vary from $-0.53^{+0.24}_{-0.21}$ up to $0.93^{+0.46}_{-0.35}$ km~s$^{-1}$~pc$^{-1}$, for Rho Oph and NGC~2237, respectively. Again, these values are typically larger than previous estimates for OB associations, but in some cases our results are less significant due to the difficulty measuring expansion in a compact cluster. Again we explored the effects of removing 2 or 3$\sigma$ outliers in position or velocity from the samples, but found that it did not have a significant effect on the results.

\begin{figure}
\centering
\includegraphics[width=240pt,trim=0 0 0 0]{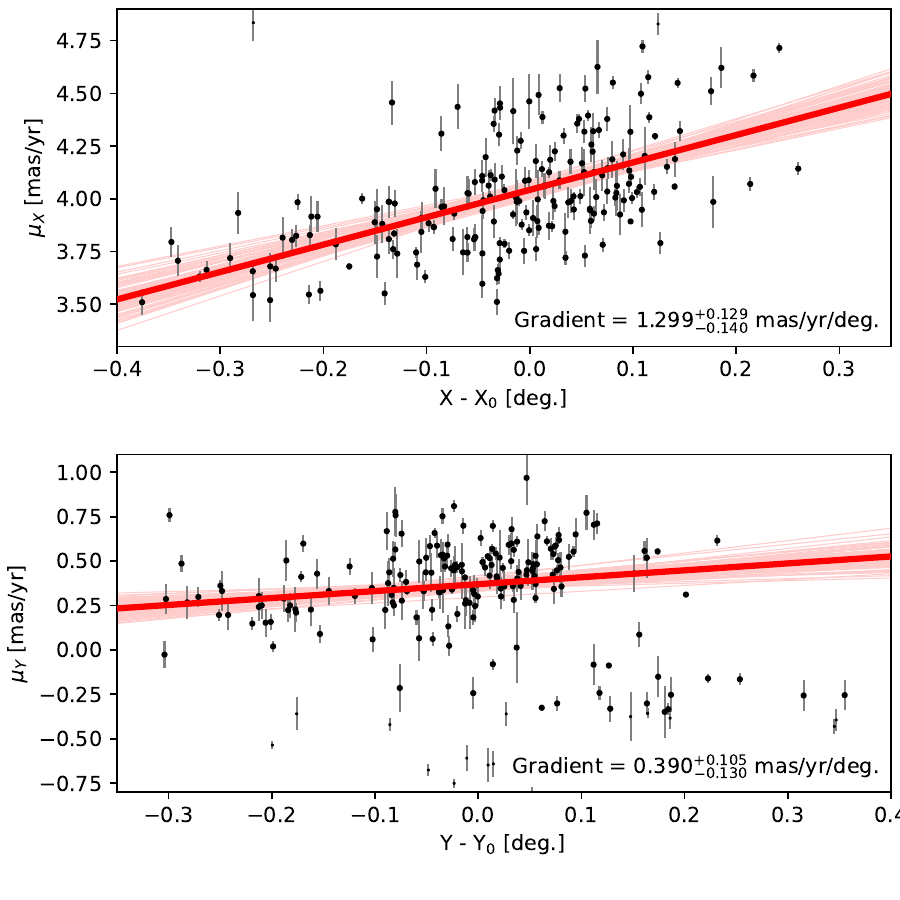}
\caption{Expansion gradients in the S~Mon cluster of NGC~2264 for 200 stars with PMs and rotated with a position angle of 120$^\circ$ (top panel) and the perpendicular axis (bottom panel). The red lines show the best-fitting expansion gradients, with 100 additional fits sampled from the posterior distribution shown in light red. The best-fitting gradient, with uncertainties are listed in each panel.}
\label{example_expansion_gradient_rotated}
\end{figure}

Figure~\ref{example_expansion_gradient_rotated} shows an example of fitted rotated expansion gradients for the S~Mon cluster of NGC~2264. These fits were performed along the axis with the strongest measured expansion gradient (with a position angle of 120$^\circ$) and the perpendicular axis. The fitted gradients are $1.30^{+0.13}_{-0.14}$ mas yr$^{-1}$ deg$^{-1}$ = $0.35^{+0.04}_{-0.04}$ km~s$^{-1}$~pc$^{-1}$ along the axis of strongest expansion and $0.39^{+0.11}_{-0.13}$ mas yr$^{-1}$ deg$^{-1}$ = $0.11^{+0.03}_{-0.04}$ km~s$^{-1}$~pc$^{-1}$ along the perpendicular axis. These $\sim$8.5 and $\sim$3 $\sigma$ measurements of expansion are notably both more significant {\it and} stronger (by approximately 50\%) than when expansion is measure along the equatorial axes.

We find that the evidence for expansion is significantly stronger when the axis of expansion is allowed to vary, as expected. In Section~\ref{s-1D_expansion} we found evidence for expansion in 11 out of 18 groups along each dimension  (albeit with many showing evidence of expansion at only the 1 $\sigma$ level), while when the axes are allowed to rotate we find evidence for expansion in 17 out of 18 groups (with only 2 at the 1 $\sigma$ level). Notable examples are $\lambda$ Ori, Barnard 35 and the S Mon cluster (all at 6$\sigma$), NGC 6530 (8$\sigma$) and ASSC 50 (13$\sigma$). The degree of expansion in the primary expansion axis is strongly correlated with the degree of expansion in the perpendicular axis, with Kendall's rank correlation test giving a correlation coefficient of 0.490 ($p$-value = 0.0039). The only group not found to be expanding is Rho Oph, which we have found to be contracting along one or more axes, regardless of the orientation of the axes.

In the groups with evidence for expansion, the level of anisotropy is broadly the same as when expansion was explored along the equatorial axes. 10 out of 17 groups show evidence for significantly ($>$2$\sigma$) anisotropic expansion, while the remaining 7 are consistent with either isotropic expansion or mildly significant expansion (at the 1$\sigma$ level).

\subsection{Outward motion}
\label{s_outward_motion}

Another method of quantifying the presence and significance of expansion is to separate the velocities of stars into their radial and azimuthal components (relative to the centre of each group) in the plane of the sky \citep[e.g.,][]{wrig16,kuhn19}. To do this we must estimate the centre of each group, which we do using the results from the EFF profile fitting (Section~\ref{s-virialstate}).

We follow \citet{kuhn19} and calculate the weighted-median outward velocity, $v_{out}$, for each group, calculating uncertainties using a Monte Carlo process that takes into account all observational uncertainties as well as the inherent uncertainties on the calculation of a median. The results can be found in Table~\ref{expansion_table}. We measure median outward velocities of up to $1.37^{+0.09}_{-0.28}$ km~s$^{-1}$ (for Trumpler 16) with most values around 0.1--0.5 km~s$^{-1}$. These values are consistent with those found by \citet{kuhn19}, albeit with smaller uncertainties, and for the five clusters in both samples (S Mon cluster, NGC 6530, NGC 2244, Trumpler 14 and Trumpler 16), four of them agree within 1$\sigma$.

The median outward velocity method gives broadly similar results to our previous methods, with 15 out of 18 groups showing evidence for expansion by this metric. Notable examples that are consistent with the method of fitting linear expansion gradients include ASSC 50 (20$\sigma$) and NGC 6530 (36$\sigma$), while notably different results are obtained, for example, for 25 Ori, which is found to be expanding with 16$\sigma$ significance by the median outward velocity method, but only 4$\sigma$ using linear expansion gradients.

As with the previous methods, Rho Oph is again found to be contracting using the median outward velocity method, at a significance of 8$\sigma$, providing further evidence of interesting kinematic behaviour in this young group.

\subsection{Summary of expansion results}

Our results show that the vast majority of groups are expanding in at least one dimension, with 17 out of 18 groups (94\%) showing evidence for expansion when the axis of expansion is allowed to rotate (Section~\ref{s_rotated_expansion}) and 15 out of 18 groups (83\%) having positive values of the median outward velocity (Section~\ref{s_outward_motion}). Only one group shows clear and consistent evidence for contraction and that is Rho Oph, which is contracting according to all our methods (with a significance of 1--8 $\sigma$ depending on the method).

More than half of all groups show evidence for expansion in at least two dimensions (11 out of 18 or 61\%), particularly when allowing the axes of expansion to rotate. The majority of groups that are expanding are doing so asymmetrically, with only 2 out of 17 of the expanding groups being consistent with symmetric expansion within 1$\sigma$, those being Cha I South (for which the uncertainties on the expansion gradients are very large) and Gamma 2 Vel (which has very low levels of expansion).

To estimate the uncertainties on the fraction of systems that are expanding (since measurement errors play a large role in the expansion gradients for some systems), we perform a Monte Carlo experiment to determine the underlying fraction of systems that are expanding. We find that the effect of measurement uncertainty is to reduce the fraction of systems observed to be expanding, particularly for the median outward velocity method. We find that, using the rotated expansion gradient fitting method, that $95^{+4}_{-6}$ \% of systems are expanding, while using the median outward velocity method that $99^{+1}_{-1}$ \% of systems are expanding.

\section{Comparison of group and kinematic properties}
\label{s-comparison}

In this section we compare the physical properties of our sample of groups (their mass, radius and age) with their kinematic properties (velocity dispersion, virial state and expansion rates) to search for correlations that might expose the physical processes at work. The values of group age and mass are included in Tables \ref{clusters} and \ref{table:virial}, respectively, compiled from the literature, many of which do not report uncertainties. In our experience, cluster ages may be inaccurate by up to 50\% and cluster masses by 20--30\%, which we have included in the fits performed here\footnote{These uncertainties are likely to be correlated between clusters as they depend on stellar evolutionary models (for ages) or the initial mass function and binary properties (for masses), each of which have standard values that are commonly used between studies, and therefore the impact of such uncertainties are greatly reduced for the rank correlation tests used in this section.}. We also calculate dynamical and relaxation timescales according to the equations in \citet{port10}. A number of obvious correlations are observed that we do not discuss in detail here, such as strong correlations between the 3D velocity dispersion and the 3D virial velocity dispersion, as well as between the virial ratio and the relaxation timescale.

\begin{figure}
\centering
\includegraphics[width=240pt,trim=0 0 0 0]{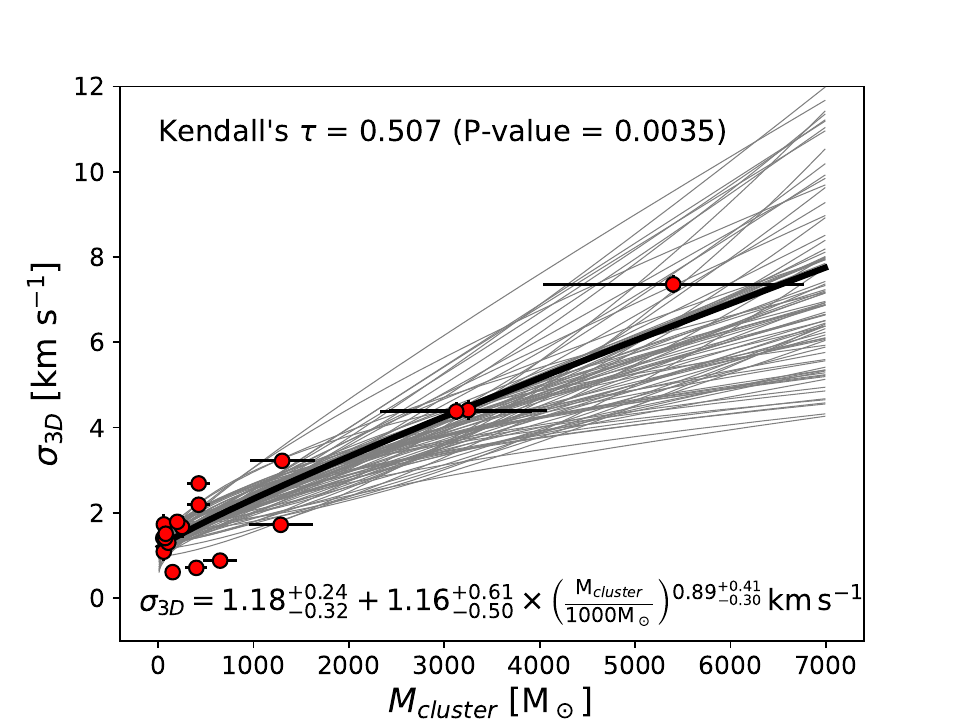}
\caption{Relationship between the 3D velocity dispersion of all systems in our sample and their total stellar mass, with uncertainties shown for both. Kendall's rank correlation test reveals a correlation of $\tau = 0.507$ and a $p$-value of 0.0035 indicating a strong positive correlation. A linear fit between the two quantities is shown in black, of the form $\sigma_{3D} = 1.18^{+0.24}_{-0.32} + 1.16^{+0.61}_{-0.50} \times (M_{cluster} / 1000 \, \mathrm{M}_{\odot} )^{0.89^{+0.41}_{-0.30}}$ km~s$^{-1}$, with grey lines showing 100 randomly sampled fits from the posterior distribution of fitted gradients.}
\label{correlation_mass_sigma}
\end{figure}

\begin{figure}
\centering
\includegraphics[width=240pt,trim=0 0 0 0]{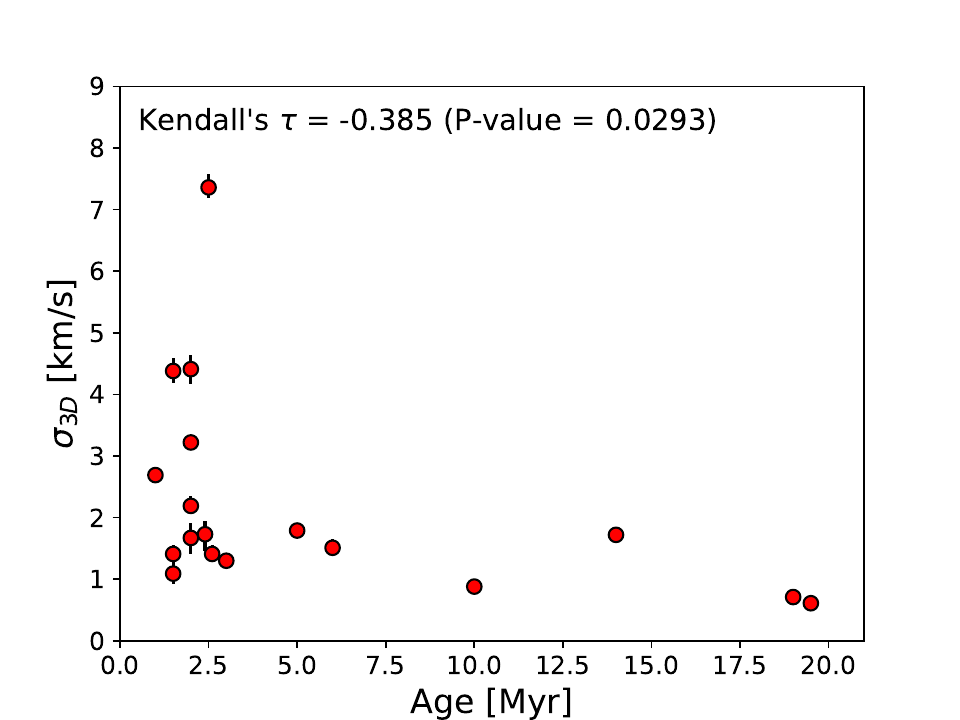}
\caption{Relationship between the 3D velocity dispersion of all systems in our sample and their age, with uncertainties shown for the former (uncertainties for the latter are likely to be highly correlated so are not shown). Kendall's rank correlation test reveals a correlation of $\tau = -0.385$ and a $p$-value of 0.0293 indicating a strong inverse correlation.}
\label{correlation_age_sigma}
\end{figure}

Figure~\ref{correlation_mass_sigma} shows the relationship between the 3D velocity dispersion and the stellar mass of the system. We find a strong correlation between these two quantities using Kendall's rank correlation test with a $p$-value of 0.0035. A similar correlation between mass and velocity dispersion was observed by \citet{kuhn19}, who also found a strong correlation between velocity dispersion and group radius, but found that since mass and radius were related (a correlation that we do not find), argued that this correlation was driven by this interdependency. A correlation between these two quantities would be expected based on the assumption of virial equilibrium, though very few of these systems were found to be in virial equilibrium. We fit a relationship of the form $\sigma_{3D} = A + B \times M_{cluster}^C$ between these two quantities using Bayesian inference and an MCMC ensemble sampler to derive a fit of

\begin{equation}
\sigma_{3D} = 1.18^{+0.24}_{-0.32} + 1.16^{+0.61}_{-0.50} \left( \frac{ M_{cluster} }{ 1000 \, \mathrm{M}_{\odot} } \right)^{0.89^{+0.41}_{-0.30} } \mathrm{km \, s}^{-1}
\end{equation}

\noindent
which is shown in Figure~\ref{correlation_mass_sigma}. The posterior distribution on the power-law index, $C$, derived from this fit gives a probability of 0.09 that the power-law index is $\le$0.5, the dependence that would be expected according to the virial relationship.

We find a strong inverse correlation between velocity dispersion and age with $p = 0.0293$ (Figure~\ref{correlation_age_sigma}). This could be due to both dynamical evolution (the most rapidly moving stars will escape a group over time, causing the measured velocity dispersion of stars within the group to reduce) and an evolutionary bias (systems with smaller velocity dispersions should survive for longer times as visible over-densities of young stars that would be selected and observed).

\begin{figure}
\centering
\includegraphics[width=240pt,trim=0 0 0 0]{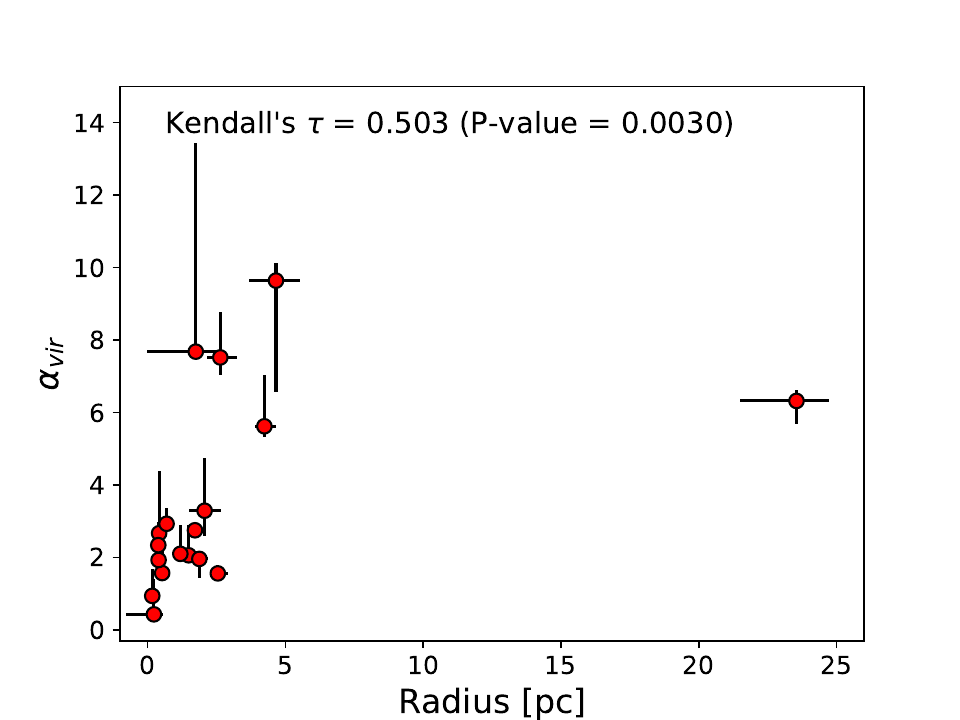}
\caption{Relationship between the virial ratio, $\alpha_{vir} = \sigma_{3D} / \sigma_{vir}$, of all systems in our sample and their radius, with uncertainties shown for both quantities. Kendall's rank correlation test reveals a correlation of $\tau = 0.516$ and a $p$-value of 0.0022 indicating a strong positive correlation.}
\label{correlation_radius_virial}
\end{figure}

\begin{figure}
\centering
\includegraphics[width=240pt,trim=0 0 0 0]{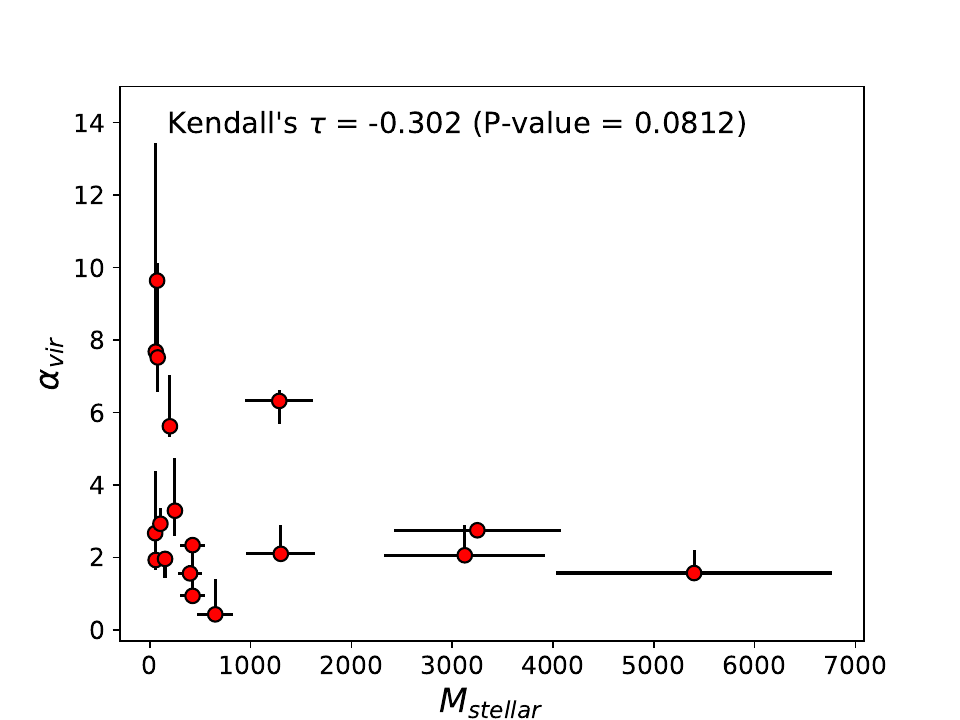}
\caption{Relationship between the virial ratio, $\alpha_{vir} = \sigma_{3D} / \sigma_{vir}$, of all systems in our sample and their stellar mass, with uncertainties shown for both. Kendall's rank correlation test reveals a correlation of $\tau = -0.302$ and a $p$-value of 0.081 indicating a weak inverse correlation.}
\label{correlation_mass_virial}
\end{figure}

We observe a strong positive correlation between the virial ratio of a system and its radius ($p$-value = 0.003), as shown in Figure~\ref{correlation_radius_virial}. The outlier in Figure~\ref{correlation_radius_virial} with a large radius is the OB association Vela~OB2 and if it is removed the correlation is stronger and suggests a linear correlation between the two quantities. There are two possible reasons for this correlation. The first is an evolutionary effect whereby gravitationally unbound systems (those with $\alpha_{vir} > 1$) will expand to larger radii, with the more unbound systems expanding faster. This should lead to a correlation between $\alpha_{vir}$ and radius, though larger systems are generally less dense and harder to detect, which might introduce a counter bias. The second reason is that $\alpha_{vir}$ is linearly dependent on the radius, and therefore if all other parameters that $\alpha_{vir}$ depends on ($\sigma_{3D}$ and $M$ specifically) either remained constant or had dependencies that cancelled each other out, one would expect to observe a linear correlation between $\alpha_{vir}$ and radius. Since this does not appear to be the case (see e.g., Figure~\ref{correlation_mass_virial}), the former evolutionary reason may be the dominant cause of this correlation.

We observe a weak inverse correlation between the virial ratio of a system and its stellar mass (Figure~\ref{correlation_mass_virial}) with a $p$-value of 0.081. However, this relationship is likely affected by an observational bias whereby more massive, unbound groups (such as massive OB associations, which would occupy the top-right of this diagram) would be larger than less massive unbound things, and therefore harder to observe by GES due to the relatively small field-of-view of the FLAMES instrument. It will be necessary to expand our sample of clusters and associations to determine the validity of this correlation.

\begin{figure}
\centering
\includegraphics[width=240pt,trim=0 0 0 0]{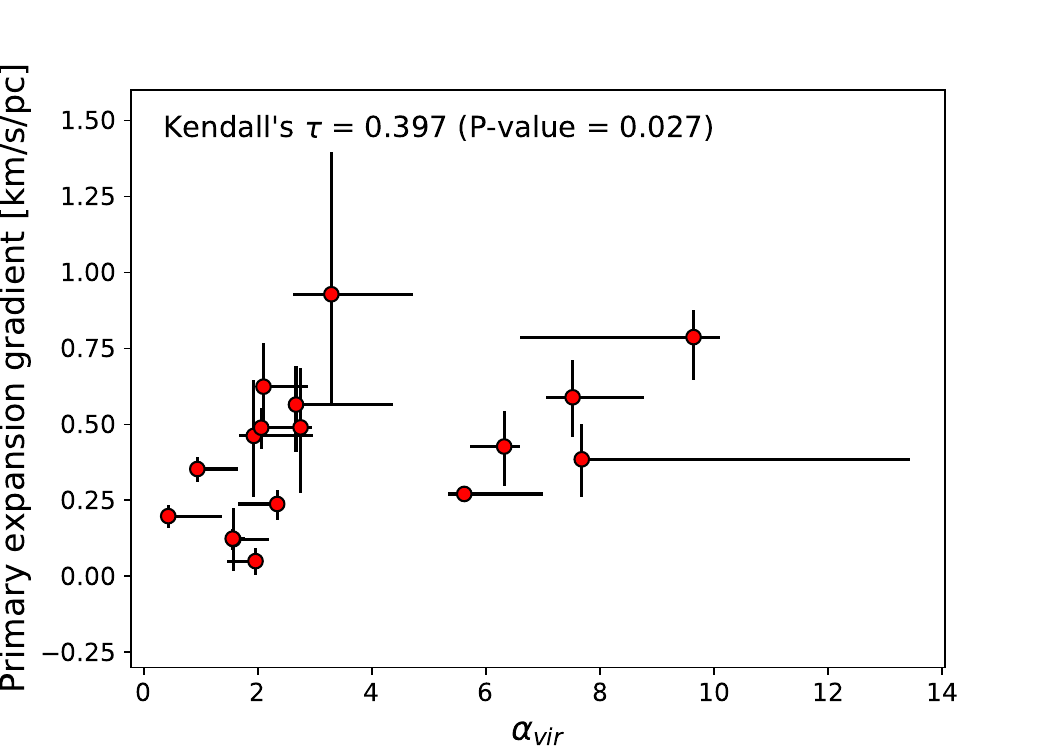}
\caption{Relationship between the virial ratio, $\alpha_{vir} = \sigma_{3D} / \sigma_{vir}$, of all systems in our sample and their expansion gradient along their primary axis of expansion (excluding the only non-expanding system, Rho Oph), with uncertainties  for both. Kendall's rank correlation test reveals a correlation of $\tau = 0.397$ and a $p$-value of 0.027 indicating a strong correlation.}
\label{correlation_virial_expansion}
\end{figure}

We observe a strong positive correlation between the virial ratio and the primary rotated expansion gradient, with a $p$-value of 0.068 (no correlation is found with any of the other measures of expansion), which improves to 0.027 when the only non-expanding system, Rho Oph, is removed (see Figure~\ref{correlation_virial_expansion}). This suggests that the more super-virial a system is (the higher its virial ratio), the higher its expansion rate will be (at least when determined by fitting rotated expansion gradients), which is consistent with a picture wherein the expansion of a group of stars is dictated by how far out of virial equilibrium they are. Notably we do not observe any correlation between the velocity dispersion and any measure of the expansion of a system (either the expansion gradients or the median outward velocity). We do however observe strong correlations between the primary and secondary expansion gradients and $v_{out}$, suggesting they are all measuring similar properties of a system.

We also observe a very strong positive correlation between the virial ratio of a system and its dynamical timescale ($p$-value = $7.0 \times 10^{-5}$). This is most likely a product of the previous correlation, in which the more super-virial a system is, the more it expands, the larger it becomes and the more its density is reduced, and therefore the longer its dynamical timescale becomes.

Finally we also observe a strong positive correlation ($p$-value = 0.027) between radius and age, with older systems being larger. Correlations such as this have been known about, in many guises, for over half a century \citep[e.g.,][]{blaa64,pfal09,kuhn15b,getm18} and are generally interpreted as due to the relaxation or expansion of systems as they age. Related to this is a strong positive correlation between age and the dynamical timescale ($p$-value = 0.014), which is likely to be a product of this since the dynamical timescale is a strong function of the group density (and therefore radius). We also observe a strong inverse correlation ($p$-value = 0.0013) between gas mass associated with the system and radius, such that the larger the system the less gas it is associated with. As larger groups are generally older, they are less likely to be associated with molecular gas (as the parental molecular cloud is either consumed or dispersed).

\section{Discussion}
\label{s-discussion}

We have performed the first large-scale 3D kinematic study of multiple stellar groups to determine and compare their structural, kinematic and evolutionary properties. We have measured 3D velocity dispersions, anisotropy levels, virial states, expansion or contraction rates and directions, as well as half-mass radii and expansion timescales. We have compared these properties with each other, as well as with literature ages and group masses, and dynamical and relaxation timescales, to identify possible correlations. In this section we discuss the meaning and implications of these results on our understanding of the dynamics, formation, evolution and dispersal of young stellar groups.

\subsection{Dynamical and virial state of young groups}

The young stellar groups studied in this work typically have 3D velocity dispersions of 1--2 km~s$^{-1}$, ranging from $0.61^{+0.03}_{-0.02}$ km~s$^{-1}$ for Gamma Vel to $7.36^{+0.19}_{-0.15}$ km~s$^{-1}$ for Trumpler 14. These results are in good agreement with previous estimates, but are typically slightly lower. This difference is likely due to our finding that simple 1D or 2D (unrotated) velocity dispersion models are likely to over-estimate the velocity dispersion by 10\% or more compared to more advanced 2D or 3D (rotated) velocity dispersion models.

We find that nearly all groups exhibit anisotropic velocity dispersions. This is generally taken to mean that the group is not sufficiently dynamically mixed to have developed an isotropic velocity dispersion. Anisotropic velocity dispersions have been observed in many regions, particularly in OB associations \citep{wrig16,wrig18} where it can be exacerbated by kinematic substructure within the association.

We find that the majority of groups are super-virial (when considering the gravitational potential due to the stellar mass), with only two groups in virial equilibrium ($\lambda$ Ori and the S Mon cluster in NGC~2264). When the mass of both stars and the surrounding molecular cloud are taken into account, all six groups with estimated molecular cloud masses in the literature are sub-virial. However, these molecular clouds are significantly more extended than their associated groups and therefore it may not always be appropriate to consider their full mass when estimating their gravitational potential.

In Section~\ref{s-comparison} we observed a positive correlation between the velocity dispersion of stellar groups and their stellar mass. Such a relation would be expected if all groups were in virial equilibrium since the virial velocity dispersion scales as $M^{0.5}$ (Equation~\ref{eqn-virial}). However, we observe and fit an approximately linear relationship between group mass and velocity dispersion with a power-law index of $0.94^{+0.42}_{-0.31}$ and were able to rule out a scaling of $M^{0.5}$ with a confidence of 93\%. Given this, the ratio of the velocity dispersion to the virial velocity dispersion would be expected to scale as

\begin{equation}
\alpha = \frac{ \sigma_{3D} }{ \sigma_{3D, vir} } \propto \frac{ M^{0.94} } { (M/R)^{0.5} } \propto (MR)^{0.5} .
\end{equation}

\noindent The mass dependence of cluster radii has been studied by various authors with dependencies varying from approximately $R \propto M^{0.25}$ \citep{brow21,dobb22} to $R \propto M^{0.5}$ \citep{adam06,pfal11}. This leads to a mass dependency of $\alpha \propto M^{0.625}$ to $M^{0.75}$, implying that less massive groups are more likely to be born in a virial or sub-virial state than more massive groups. When comparing the virial ratio with the group mass, we were unable to identify a correspondingly strong correlation due possibly to selection biases.

\subsection{Expanding star clusters and groups}

We find that the vast majority of groups show evidence that they are expanding. The exact fraction depends on the method used to measure expansion, varying from 83\% using the median outward velocity, 89\% using 1D expansion gradients, and 94\% using 2D rotated expansion gradient fits. The significance of these results vary, but 12 of the 17 groups with evidence for expansion (71\%) do so at the 3$\sigma$ level. The strongest expansion is found for the groups NGC~2237 and Barnard 35, which are expanding at rates of $0.93^{+0.46}_{-0.35}$ and $0.79^{+0.08}_{-0.14}$ km~s$^{-1}$~pc$^{-1}$ (at significances of 2.7 and 5.7$\sigma$), as determined using 2D rotated expansion gradient fits.

We find that a larger fraction of groups are expanding compared to previous studies. \citet{kuhn19} found that 75\% of their groups had positive median outward velocities (compared to 83\% here), but that only 57\% of their groups had positive median outward velocities to a confidence of $>$1$\sigma$ (compared to 78\% here). In studies of larger OB associations, many historical studies have struggled to identify large-scale expansion patterns \citep[e.g.,][]{wrig16,wrig18}, but more recent studies that dissected OB associations into subgroups using kinematic data have almost universally been able to identify expansion \citep[e.g.,][]{cant19b,arms22,quin23}. Our results support the view that the subgroups of OB associations represent the expanded remnants of compact groups similar to the clusters studied in this work and that OB associations are therefore composed of multiple expanding star clusters \citep[e.g.,][]{wrig22}.

The majority of groups that are expanding are doing so anisotropically, i.e., with different gradients along different axes. This is the case whether the expansion gradients are fitted in 1D or 2D. This implies that, if no other forces are acting on the stars during their expansion, that these groups either started their expansion from non-spherical initial conditions or had anisotropic velocity dispersions prior to expansion. This is consistent with previous studies that have found that the majority of young groups are non-spherical \citep[e.g.,][]{kuhn14} and our result that the majority of groups have anisotropic velocity dispersions. Gravitational tidal forces from the surrounding molecular cloud \citep{krui12b} or the residual molecular gas that has since been dispersed by feedback \citep{zamo19} could both lead to asymmetric expansion of the group if the gas is non-spherically distributed.

There are two groups in our sample that are in virial equilibrium but observed to be expanding; the S Mon cluster of NGC 2264 ($\alpha_{vir} = 0.94^{+0.68}_{-0.01}$) and $\lambda$ Ori ($\alpha_{vir} = 0.43^{+0.92}_{-0.01}$). If these systems are in virial equilibrium then these results would appear to be counter-intuitive. However, we note that both of these groups have (primary) rotated expansion gradients at the lower end of those measured (respectively 0.35 and 0.20 km~s$^{-1}$~pc$^{-1}$). This is part of a strong correlation between the virial ratio and the (primary) rotated expansion gradient that has a $p$-value of 0.034. If these systems are in virial equilibrium then their observed expansion may simply be due to the system settling down into an equilibrium configuration following formation \citep{park16,sill18}.

\subsection{Expansion timescales and ages}

Expansion timescales can be calculated from the expansion gradients by simple inversion (with a correction factor to account for  units). We calculate expansion timescales for all our groups using the largest expansion rate (from the rotated expansion gradient fit) as this will give the smallest expansion timescale for the two expansion axes. We find that the expansion timescales vary from $\sim$1 Myr for NGC~2237 and Barnard 35, to $\sim$8 Myr for 25 Ori and Trumpler 14, and $\sim$20 Myr for Gamma Vel (though the expansion rates for Trumpler 14 and Gamma Vel are of marginal significance).

The majority of groups have expansion timescales that are very similar to their evolutionary ages, though there are some outliers that mean the measured correlation is very weak with a $p$-value of only 0.25. Multiple factors can affect the agreement between expansion timescale and age of the system, including delayed expansion (the group not expanding immediately after star formation), non-compact initial conditions (the group not expanding from an initially compact configuration), or additional forces acting on the stars \citep[which may act to either accelerate or decelerate the expansion, e.g.,][]{zamo19}. These factors can each cause the expansion age to be over- or under-estimated relative to the true age of the system.

There are three systems with expansion timescales that are significantly smaller than their evolutionary ages; Vela OB2 (2.3 vs 14 Myr), $\lambda$ Ori (4.9 vs 10 Myr) and Col. 197 (1.6 vs 5 Myr). Vela OB2 is an OB association with considerable kinematic substructure \citep[e.g.,][]{cant19b,arms22} that would not have expanded from initially compact initial conditions, which goes a long way towards explaining this disagreement. $\lambda$ Ori has an expansion timescale of $\sim$5 Myr, significantly smaller than its evolutionary age of $\sim$10~Myr from \citet{bell13}, but there is some disagreement in the age of this system as \citet{koun18} estimate an evolutionary age of 4--5~Myr, which would be in better agreement with the expansion age. Finally, the expansion timescale of $\sim$1.6~Myr for Col.~197 is in sharp contrast to its evolutionary age of $\sim$5~Myr. Since the group is embedded within an H~{\sc ii} region (Gum 15) this might indicate that the group is younger than previously thought, or alternatively that the expansion of the system did not begin immediately after formation.

There are two systems with expansion timescales that are significantly larger than their evolutionary ages; Trumpler 14 (8 vs 2.5 Myr) and the Spokes cluster within NGC~2264 (4.1 vs 1 Myr). The former has large uncertainties on its evolutionary age, so this is not significant, but the latter is more significant. The most likely explanation for this is that this cluster did not begin its expansion from highly compact initial conditions, but rather did so from conditions more similar to how it is currently observed and has been slowly expanding since formation.

\subsection{Contracting systems: Rho Ophiuchus}

There is one system in our sample that is contracting rather than expanding: Rho Ophiuchus. Some systems exhibit negative expansion gradients along one axis, negative 1D expansion gradients or negative median outward velocities, but for nearly all of these systems these measurements are either insignificant or give different results depending on how the expansion is measured.  Rho Oph however appears to be contracting according to all three methods used to measure expansion and contraction, with approximately 2.5$\sigma$ significance using the 1D or 2D expansion gradient fits, and with $\sim$4$\sigma$ significance using the median outward velocity method. This implies that Rho Oph is almost definitely contracting.

The virial ratio of Rho Oph, when considering only stars, is $\alpha_{vir} = 2.93^{+0.38}_{-0.13}$, suggesting that the system is gravitationally unbound. However, when one takes into account the approximately 1750~M$_\odot$ \citep{lore89} of gas in the molecular cloud the virial ratio drops to $0.70^{+0.09}_{-0.03}$ suggesting that the group is gravitationally bound, and in particular is sub-virial. \citet{rigl16} came to a similar conclusion regarding the virial state of Rho Oph comparing their velocity dispersion to the mass of the surrounding molecular cloud and estimated an 80\% probability that the group is gravitationally bound. The young stars observed by GES are part of the optically visible 3~Myr \citep{gras21} population that surrounds (on the plane of the sky) the L1688 molecular cloud and so presumably would feel the gravitational pull of its mass. The molecular cloud could therefore be responsible for the contraction of the surrounding group of young stars. This may lead to the accretion of these young stars onto the group of forming stars within L1688. If so, this would provide some of the first direct kinematic evidence of a group accreting other, already formed, young stars.

Simulations of star cluster formation show that many systems undergo an initial collapse during the first crossing time of the system \citep{pros09,park16}, that could be observed as contraction. This collapse potentially leads to mergers between sub-groups \citep{bonn03,vazq17}. Observational support for the model of star cluster formation through mergers has been mixed however. \citet{park16} and \citet{arno22} identified possible kinematic signatures of past subgroup mergers that could be used to determine if mergers have taken place after the fact, but so far only NGC~6530 has been observed to exhibit any such signature \citep{wrig19b}. In their study of the dynamics of young groups, \citet{kuhn19} identified the group M17 as a system potentially undergoing collapse and mergers due to its negative median outward velocity and highly clumpy structure. However, they found no evidence for the converging motion of subgroups towards other subgroups in their sample, suggesting that if mergers take place they do so predominantly at early ages. It is therefore even more notable that Rho Ophiuchus is in the process of contracting given its estimated age of $\sim$3~Myr.

\subsection{The formation of star clusters}

\begin{figure}
\centering
\includegraphics[width=240pt,trim=0 0 0 0]{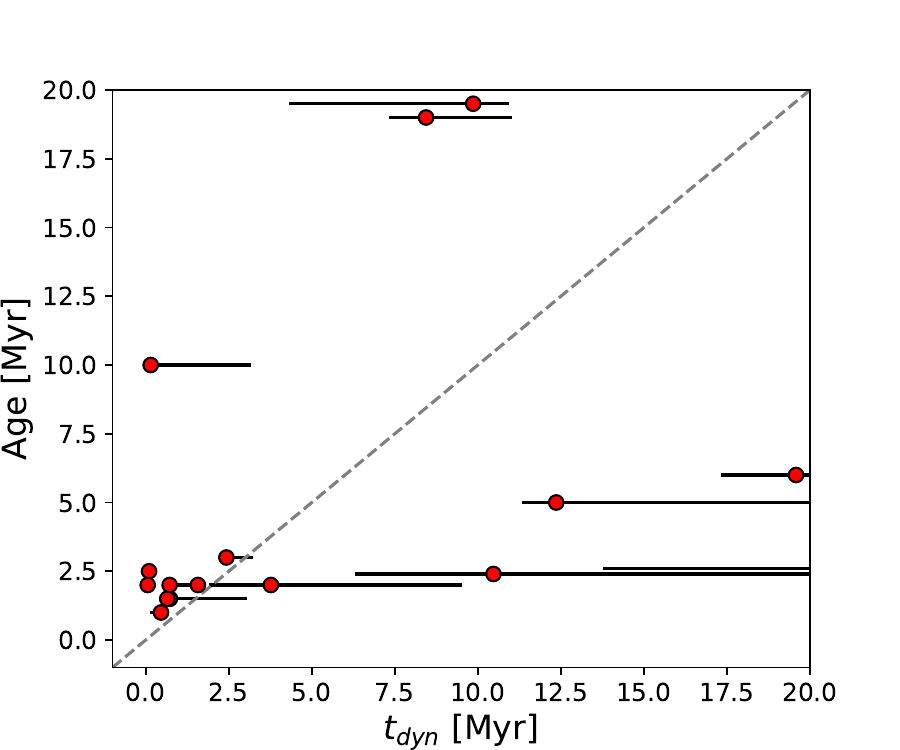}
\caption{Relationship between the dynamical timescale, $t_{dyn} = (GM/r^3_{vir})^{-1/2}$, of all systems in our sample and their literature age, with uncertainties for the former. Vela OB2 (which has an age of 14~Myr and a dynamical timescale of $\sim$120~Myr) and Barnard 35 (age 2.6~Myr and dynamical timescale $\sim$45~Myr) are not shown. A 1:1 relationship is shown as a grey dashed line.}
\label{correlation_age_dynamical_timescale}
\end{figure}

How young and compact star clusters form is still an open question, with competing theories suggesting they either form monolithically in a dense, compact distribution \citep[e.g.,][]{bane15} or that they form over a wider area and collapse down to form a compact cluster \citep[e.g.,][]{bonn03,arno24}. Studies suggest that the question of how bound star clusters form is related to both the star formation efficiency \citep{krui12} and the collapse of the giant molecular cloud \citep{chev22}.

If star clusters form monolithically then they should have existed in more-or-less their current configuration since birth, and if they are sufficiently dense then they will have had time to dynamically evolve on a timescale set by their dynamical timescale. However, the majority of groups we have studied are not dynamically evolved (lacking for example, isotropic velocity distributions), suggesting that they haven't existed in a grouped configuration for very long.

We compare the age of all the groups in this study with their dynamical timescale in Figure~\ref{correlation_age_dynamical_timescale}. This figure shows that the majority of groups (12 out of 18) are older than their dynamical timescale. The groups that are younger than their dynamical timescale (those to the right of the 1:1 line) are predominantly OB associations (Vela OB2 and ASSC 50) or very low-mass groups (Barnard 30 and 35), though it also includes NGC~2237 and Col. 197. If we exclude the known OB associations from this comparison and focus only on the known clusters or cluster candidates then the fraction increases to 12 out of 16, or 75\%.

This suggests that these systems have not existed in their current configuration since birth, otherwise they would have had time to dynamically evolve and establish some degree of isotropy. This argues against these clusters having formed monolithically and instead favours a picture whereby these clusters formed at a lower density (with a longer dynamical timescale) and have since collapsed down to form more compact clusters.

Further evidence in favour of the star cluster formation model of collapse and mergers is that all of the systems that are associated with a molecular cloud for which masses are available in the literature are in a subvirial state once the mass of the surrounding molecular cloud is taken into consideration. This shows that in these cases there is sufficient mass in the molecular cloud to introduce a state of contraction in these groups, which could then lead to the formation of a compact star cluster. This picture is further bolstered by the observation that Rho Oph is currently in a state of contraction.

\subsection{The dispersal of young star clusters and the survival of long-lived open clusters}

While the clustering of very young stars is an almost ubiquitous phenomenon, most stars do not find themselves in bound star clusters by an age of 10 Myr or older \citep{lada03}. The explanation for this is believed to be a combination of the unbinding of gravitationally-bound (embedded) clusters by residual gas expulsion\citep[e.g.,][]{hill80} and the fact that most groups of clustered stars in star forming regions are actually gravitationally unbound. Our study has provided the strongest evidence to date that the majority of young groups are in the process of expanding and do so from an age as young as $\sim$1--2 Myr. Our finding that $>$90\% of young groups are in the process of expanding is consistent with the observation by \citet{lada03} that 90\% of young `clusters' disperse within 10~Myr.

Given this, what determines which clusters will go to become long-lived open clusters? \citet{krum20} suggest that the clusters that do survive their early evolution and go on to become gravitationally bound are distinct kinematically, with isothermal, spherically symmetric density distributions, virialised velocity distributions and are neither expanding or contracting. Our study has very few groups that meet these criteria. The fraction of groups that are out of virial equilibrium (when only considering the stellar part of the system) is 94\% (only two groups are consistent with being in virial equilibirum). The fraction of groups that are inconsistent with having isotropic velocity distributions is $>$90\% (depending on the method used to measure the velocity dispersion), with only the Gamma Vel cluster and NGC~6530 being consistent with isotropy. And finally, the vast majority of groups show significant evidence for expansion or contraction, with only Gamma Vel and Trumpler 14 being consistent with not being in a state of expansion or contraction.

We conclude that the vast majority of the systems studied will expand and disperse within the next 10--20 Myr. Even the six groups that are in virial equilibrium when considering the surrounding molecular cloud are likely to disperse once they spatially decouple from the gas (including Rho Oph, despite its currently contracting state). $\lambda$ Ori may survive as a long-lived open cluster since it appears to be in virial equilibrium, though it is in a state of weak expansion (this may represent the system settling down into a stable state following formation). Both Gamma Vel and Trumpler 14 are close to being in virial equilibrium and do not exhibit strong expansion. If their stellar masses have been under-estimated or their radii over-estimated then they may be in virial equilibrium. They appear to be the two best candidates for becoming long-lived open clusters from the sample of groups studied here.

\section{Conclusions}

We have presented the first large-scale 3D kinematic study of multiple young star clusters and OB associations. The combination of {\it Gaia} astrometry with {\it Gaia}-ESO Survey spectroscopy provides 3D positions and kinematics, as well as kinematically-unbiased indicators of youth, for $\sim$2700 young stars in 18 clusters or associations. 

We measure 3D velocity dispersions for all 18 groups that range from 0.62 to 7.5 km~s$^{-1}$ (1D velocity dispersions range from 0.36 to 4.3 km~s$^{-1}$). We find that the majority of groups have anisotropic velocity dispersions, suggesting they are not  dynamically relaxed. From the 3D velocity dispersions, measured radii and estimates of total mass in the literature we determine the virial state of all groups and find that all but two systems are super-virial when only the stellar mass is considered, but that some systems are in virial equilibrium when the mass of the surrounding molecular cloud is taken into account. We observe an approximately linear correlation between the 3D velocity dispersion and the group mass, implying that the virial state of groups should scale $M^{0.625}$ -- $M^{0.75}$. However, we do not observe a strong correlation between virial state and group mass.

In agreement with their virial state we find that nearly all of the groups are in the process of expanding, as indicated by both linear expansion models and the median outward velocity of stars in the group. Given their viral state and expansion, these systems are expected to continue to expand and form OB associations (or subgroups within OB associations). In the majority of cases the expansion is anisotropic, implying that either groups are not spherical or have anisotropic velocity dispersions prior to expansion, or that additional forces act on the group during their expansion. Given that most groups are not dynamically relaxed and that other observations find many young groups to have substantial levels of ellipticity, the former explanation is argued to play at least a contributing role.

One group, Rho Oph, is found to be contracting using all measures of expansion/contraction. The group is in a super-virial state when only the stellar mass is considered, but is sub-virial when the mass of its molecular cloud is considered. Whether or not the group is currently gravitationally bound, it is currently contracting, which may lead to mergers between subgroups or the accretion of stars onto a central cluster.

We conclude that, since the majority of clusters are not dynamically evolved, despite being older than their dynamical timescales, that these clusters did not form as we observe them now, but originally were larger and had lower densities (and thus had longer dynamical timescales). Combined with other evidence, we conclude that most clusters form through the collapse of an extended distribution of stars, with mergers between subgroups, and did not form monolithically.

We also conclude that the majority of the groups studied here will not survive as long-lived open clusters, being super-virial, having non-isotropic velocity dispersions and showing significant evidence for expansion. We conclude that they will expand and disperse into the field population of our galaxy. The best candidates for survival as long-lived open clusters are Gamma Vel, Trumpler 14 and $\lambda$ Ori, each of which are either in or close to virial equilibrium.

The data and sample presented here provide a powerful illustration of the scientific potential that will arise from the combination of {\it Gaia} and data from the next generation of multi-object spectroscopic surveys such as WEAVE, 4MOST and SDSS-V. Spectroscopy from these surveys for tens to hundreds of thousands of young stars will allow this work to be extended, better sampling individual groups while also targeting a larger number of groups with a greater range of ages, masses and densities. This will help address a range of outstanding issues that this work has probed including how star clusters form, the dynamical processes at work within them, and the physical processes that drive their disruption and dissolution.

\section*{Acknowledgments}

We are grateful to N\'uria Miret-Roig, Alexis Quintana, and Steve Stahler, for reading this manuscript and sending valuable comments. NJW acknowledges an STFC Ernest Rutherford Fellowship (grant number ST/M005569/1) and a Leverhulme Trust Research Project Grant (RPG-2019-379). This research has made use of NASA's Astrophysics Data System and the Simbad and VizieR databases, operated at CDS, Strasbourg.

This work is based on data products from observations made with ESO Telescopes at the La Silla Paranal Observatory under program ID 188.B-3002. These data products have been processed by the Cambridge Astronomy Survey Unit (CASU) at the Institute of Astronomy, University of Cambridge, and by the FLAMES-UVES reduction team at INAF-Osservatorio Astrofisico di Arcetri. These data have been obtained from the Gaia-ESO Survey Data Archive, prepared and hosted by the Wide Field Astronomy Unit, Institute for Astronomy, University of Edinburgh, which is funded by the UK Science and Technology Facilities Council. This work also made use of results from the European Space Agency (ESA) space mission Gaia. Gaia data are being processed by the Gaia DPAC. Funding for the DPAC is provided by national institutions, in particular the institutions participating in the Gaia MultiLateral Agreement (MLA). The Gaia mission website is https://www.cosmos.esa.int/gaia. The Gaia archive website is https://archives.esac.esa.int/gaia.

\section*{Data Availability}

The GES and {\it Gaia} data used in this article are available from the ESO and Gaia archives, respectively. The sample used in this article and the physical quantities calculated and derived in this article will be uploaded to Vizier.

\bibliographystyle{mn2e}
\bibliography{/Users/nwright/Documents/Work/tex_papers/bibliography.bib}
\bsp

\appendix

\section{The observed groups}
\label{s_targets}

Here we summarise the information, from the literature, on the 18 young clusters, associations and star-forming regions observed by the {\it Gaia}-ESO Survey that we have studied in this work. We also briefly summarise how the targeted stars in the GES observations were separated into groups.

\begin{itemize}
\item \textbf{Rho Ophiuchus} is our closest target ($d = 136$~pc), and one of the closest star-forming regions. The entire Ophiuchus cloud complex spans $\sim 6 \times 3$ degrees on the sky ($14 \times 7$ pc), though the nine FLAMES pointings obtained by GES cover an area of only $1.4 \times 1.4$ degrees ($3.4 \times 3.4$~pc) centred on the dense molecular cloud L1688, but focussed on the optically-visible non-embedded stars \citep{rigl16}. The total optically-visible mass of stars towards L1688 is $\sim$85~M$_\odot$ \citep{eric11,rigl16}, which rises to 106~M$_\odot$ to account for unresolved binaries\footnote{Assuming a 50\% binary fraction and a mean mass ratio of 0.5, a reasonable approximation across multiple primary star masses \citep[e.g.,][]{duch13}.} with an additional $\sim$1750~M$_\odot$ in gas in L1688 \citep{rigl16}. The young stars studied here are part of Population 1 from \citet{gras21} with an average age of $\sim$3~Myr. The GES data for Rho Oph were previously studied by \citet{rigl16}.
\item \textbf{Chamaeleon I (North and South)}, our most reddened targets with $A_V \sim 3$ \citep{luhm07} and at distances of 191 and 187 pc, respectively, are also the youngest targets with an age of $\sim$1.5~Myr \citep{gall21}. Most of the known members are confined to an area of $\sim 0.7 \times 1.7$ degrees in the form of two groups, Cha I North and South \citep{gall21}, each with a radius of 0.35$^\circ$ (1.1~pc). The total number of primary stars in Cha~I has been estimated to be 226 \citep{luhm07}, which for a mean stellar mass of 0.4 M$_\odot$ and accounting for unresolved binaries equates to $\sim$113 M$_\odot$. Cha I North has $\sim$10\% fewer stars than Cha I South \citep{gall21} and so we estimate their total stellar masses as 54 and 59 M$_\odot$, respectively. The mass of the molecular cloud in which Cha~I is embedded is estimated to be 1000 M$_\odot$ \citep{mizu01}. We divide the data for Cha~I into the North ($\delta > -77^\circ$) and South ($\delta < -77^\circ$) clusters \citep{rocc18}. The GES data for Cha~I were previously studied by \citet{sacc17}.
\item \textbf{Gamma Vel and Vela OB2}, aged $\sim$ 19.5 and 14 Myr respectively \citep{jeff17,arms22}, are our oldest targets \citep[respectively Populations A and B from][]{jeff14}. Gamma Vel is a small, non-embedded cluster towards the Vela OB2 association \citep{jeff14,arms20}, lying at distances of 334 and 367 pc, respectively. The cluster has a radius of approximately 1.37~pc \citep{jeff14}, while the association spans 50--100 pc \citep{arms22}. The total mass of stars is estimated to be $\sim$152 M$_\odot$ for Gamma Vel \citep{jeff14} and $1285 \pm 110$ M$_\odot$ for Vela OB2 \citep{arms18}. We follow \citet{arms20} and select stars having $(\mu_\alpha + 6.53)^2 + (\mu_\delta - 9.8)^2 \leq 0.7^2$ mas/yr as being members of Gamma Vel, while all other stars are considered part of Vela OB2. The GES data for Gamma Vel and Vela OB2 were previously studied by \citet{jeff14} and \citet{fran18}.
\item \textbf{25 Ori} is a young, non-embedded group in the Orion OB1 association, which \citet{fran22} recently determined an age of $19^{+1.5}_{-7.0}$~Myr for. It lies at a distance of $\sim$339 pc and has a radius of approximately 0.62$^\circ$ or 3.7~pc \citep{khar05}. The mass of primary stars within a 1$^\circ$ radius is estimated to be $324 \pm 25$ M$_\odot$ \citep{suar19}, and when accounting for unresolved binaries this equates to $400 \pm 30$ M$_\odot$. We exclude stars outside of the central group as non-members (those with $\delta < 1.2^\circ$). The GES data for 25~Ori has not previously been studied dynamically.
\item \textbf{$\lambda$ Ori} is a young \citep[$\sim$10~Myr,][]{dola02}, non-embedded group in the Orion OB1 association at a distance of $\sim$389 pc. The total mass of primary stars was estimated by \citet{dola02} to be 450--650 M$_\odot$. When accounting for unresolved binaries and revising for the nearer distance from {\it Gaia} we estimate the total stellar mass to be $650 \pm 120$ M$_\odot$. The group radius is approximately 3$^\circ$ or $\sim$20~pc. The GES data for $\lambda$ Ori has not previously been studied dynamically.
\item \textbf{Barnard 30 \& Barnard 35} are two dark clouds in the vicinity of $\lambda$ Ori (north-west and south-east of $\lambda$ Ori respectively) at similar distances (384 and 390 pc respectively), and both containing young stellar groups. They were observed as part of the $\lambda$ Ori observations. Their ages are estimated to be $2.4 \pm 1.3$ and $2.6 \pm 1.3$ Myr \citep{koun18}, notably younger than the nearby $\lambda$ Ori group. Their total stellar or gas masses are not precisely known, but \citet{koun18} identify 96 and 117 optically-exposed members, respectively, with an estimated completeness of 80\%. Combining this with a mean stellar mass of 0.4~M$_\odot$ and accounting for binaries gives estimated stellar masses of 60 and 73 M$_\odot$, respectively, for Barnard 30 and 35. No estimates of the gas or dust mass in either dark cloud could be found in the literature. The GES data for Barnard 30 and 35 have not previously been studied dynamically.
\item \textbf{The S Mon Cluster and the Spokes Cluster (NGC 2264)}, are two young clusters in the NGC 2264 H~{\sc ii} region. NGC~2264 is a highly substructured region elongated along a $\sim$10~pc NW-SE orientation. The Spokes cluster (sometimes called NGC 2264-C) is centred around the Class I protostar IRS2 in the south \citep{teix06}, while the S Mon cluster (sometimes referred to as the Christmas Tree cluster) is associated with the O7 binary S Mon \citep{sung09} in the north. Both clusters are young, with ages of $\sim$1 Myr for the Spokes cluster and $\sim$2 Myr for the S Mon cluster \citep{venu19}, and partially embedded. They each have approximate sizes of $\sim$1~pc in diameter. The total stellar population in NGC 2264 is estimated to be $\sim$1700 members, approximately equally divided between the two clusters \citep{venu19}. This equates to a total mass of each cluster of 425 M$_\odot$, when accounting for unresolved binaries. \citet{oliv96b} estimated the molecular cloud that NGC~2264 is embedded within has a total mass of $\sim$28000 M$_\odot$ (when scaled to our {\it Gaia} distance), but from molecular maps we estimate that only $\sim$10\% of the molecular cloud's mass ($\sim$3000~M$_\odot$) is centred around each cluster and therefore relevant for its dynamics. We select members of the Spokes cluster as those with $\delta < 9.72^\circ$ and members of the S Mon cluster with $\delta > 9.72^\circ$. The GES data for NGC~2264 has not previously been studied dynamically.
\item \textbf{ASCC 50 (Alessi 43)} is a young group in the Vela T2 association \citep{pett94}. The group was first detected by \citet{khar05} in the RCW 33 H~{\sc ii} region and has a {\it Gaia} distance of $912 \pm 3$~pc. The age of the group has been estimated from $\sim$5 Myrs \citep{pris18} to $\sim$11.5 Myrs \citep{cant20}. The most massive member is the O9V+B0V binary HD~75759, which would have a mass of 17--20 M$_\odot$ and if on the main sequence an age $\lesssim 5-6$ Myr \citep{ekst12}, therefore we adopt an age estimate of $\sim$5 Myr. If this were the most massive star in the group it would imply a group mass of 300-400 M$_\odot$ \citep{weid06}. \citet{pris18} estimate a total mass of 50--86 M$_\odot$, which when scaled to account for binarity gives a mass of $85 \pm 23$ M$_\odot$. We compromise and use a mass of $200 \pm 100$ M$_\odot$. The group has a radius of $\sim$0.25 degrees, or 4 pc at a distance of 912 pc. The GES data for ASCC 50 have not previously been studied dynamically.
\item \textbf{Collinder 197} is a group of young \citep[$\sim$6 Myr,][]{pris18} stars at a distance of 925 pc in Vela. The total mass of the group was estimated by \citet{bona10} to be $660^{+102}_{-59}$ M$_\odot$, though \citet{pris18} estimate a mass of 45--81 M$_\odot$, or $79 \pm 23$ M$_\odot$ when accounting for binaries. The latter seems more consistent with the group's most massive member being the B3/5II star HD~74804 \citep[that we estimate has a mass of 5.5~M$_\odot$ from fitting its photometry to evolutionary models,][]{ekst12}, and therefore we use that mass here. The radius of the group is approximately 12 pc \citep{bona10}. The GES data for Collinder 197 has not previously been studied dynamically.
\item \textbf{NGC 6530 (the Lagoon Nebula)}, is the group with the most GES spectra, with $\sim$650 members identified. The group is $\sim$1.5 Myr old \citep{bell13}, with a radius of $\sim$2 pc \citep{wrig19} at a distance of 1.24 kpc. The total stellar mass of NGC 6530, including binaries, has been estimated to be 3125 M$_\odot$ \citep{wrig19}, while the molecular cloud it is associated with has an estimated mass of 40,000 M$_\odot$ \citep{take10}. The GES data for NGC 6530 was previously studied by \citet{wrig19} and \citet{wrig19b}.
\item \textbf{NGC 2244 (the Rosette Nebula)} is a 2 Myr \citep{bell13} rich cluster containing $>$70 OB stars \citep{wang08}. The total stellar mass has not been well measured in the literature, but is approximately 1300 M$_\odot$ based on a mass of 40--50 M$_\odot$ for its most massive member, the O4V star HD 46223. Such a mass would be consistent with suggestions that the cluster contains $\sim$2000 young stars \citep{wang08}. The cluster has a radius of approximately 4 pc \citep{wang08} at a distance of $\sim$1.37 kpc. The entire Rosette Molecular Cloud has been estimated to have a total mass of $\sim$$10^5$ M$_\odot$ \citep{blit80}, though NGC~2244 is not embedded within the cloud, which actually surrounds the cluster. The GES data for NGC 2244 have not previously been studied dynamically.
\item \textbf{NGC 2237} is a young star cluster projected against the periphery of the Rosette Nebula, but $\sim$130 pc behind it (at a distance of 1.49 kpc). It was observed as part of the GES observations of NGC~2244.  It is estimated to be $\sim$2 Myr old \citep{wang10} and contain 400--600 stars \citep{wang10}, suggesting a total mass, including binaries, of $\sim$250 M$_\odot$ (though note that both age and mass estimates were derived on the assumption that NGC 2237 was at the same distance as NGC 2244, 10\% closer than its {\it Gaia} EDR3 parallax implies). Its radius is $\sim$0.2 degrees or $\sim$5 pc at 1.49 kpc. The GES data for NGC 2237 have not previously been studied dynamically.
\item \textbf{Trumpler 14 \& 16}, our most distant targets ($d \sim 2.2$~kpc), are both young \citep[2.5 and 2 Myr respectively,][]{hur12}, compact \citep[radii $\sim 1$~pc,][]{asce07}, but not centrally concentrated \citep{reit19}. Trumpler 14 has an approximate total mass of $5400^{+4100}_{-1900}$ M$_\odot$ \citep[][once unresolved binaries are accounted for]{sana10}. Trumpler 16 is comparable, but slightly less massive. \citet{wolk11} estimates a total stellar population of $6500 \pm 650$ stars, which suggests a total mass, including binaries, of $3250 \pm 325$ M$_\odot$. We separate the two clusters spatially, with Trumpler 14 members having $\delta > -59.76^\circ + 0.35 \, (\alpha - 160.8^\circ)$. The GES data for both Trumpler 14 and Trumpler 16 have not previously been studied dynamically.
\end{itemize}

\section{Spectroscopic membership indicators}

\begin{figure*} 
\centering
\includegraphics[width=14.5cm,trim=100 110 100 340]{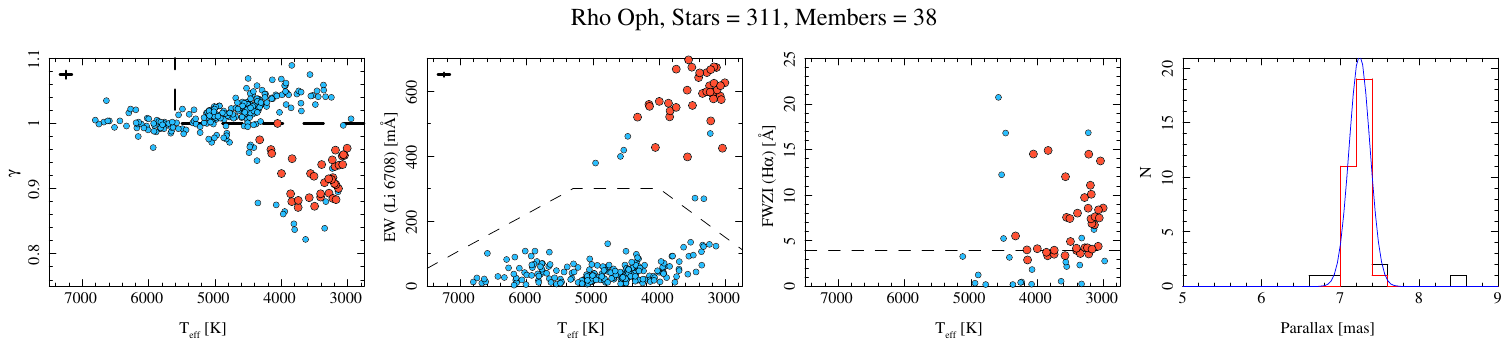}
\includegraphics[width=14.5cm,trim=100 110 100 340]{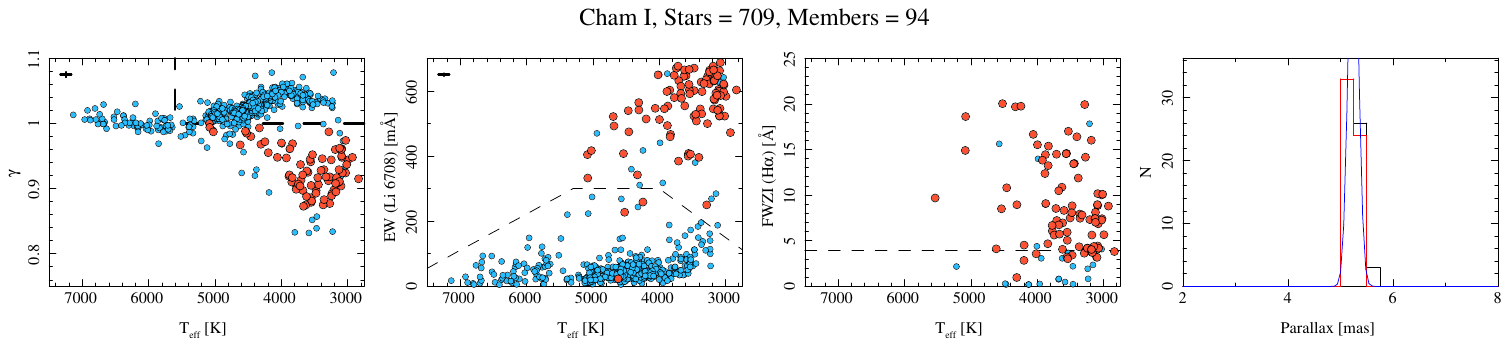}
\includegraphics[width=14.5cm,trim=100 110 100 340]{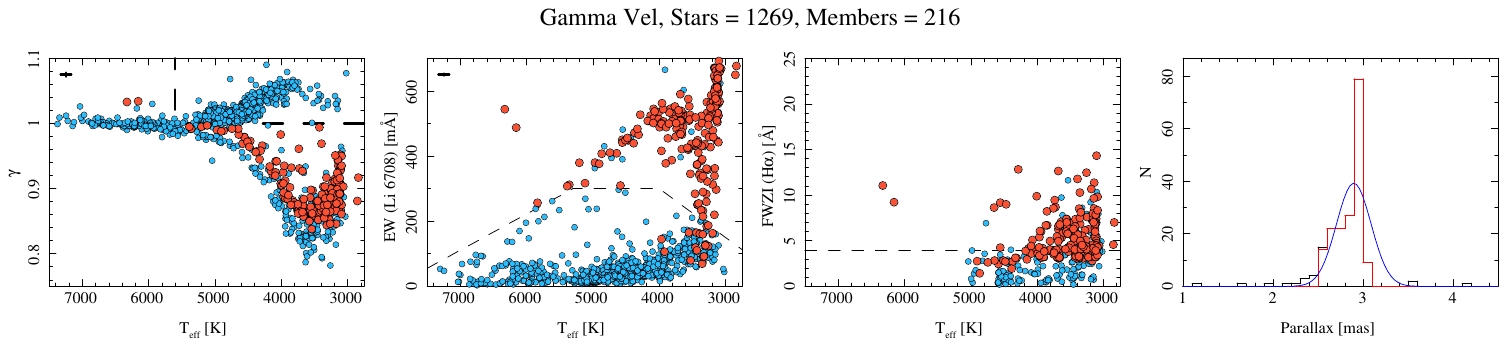}
\includegraphics[width=14.5cm,trim=100 110 100 340]{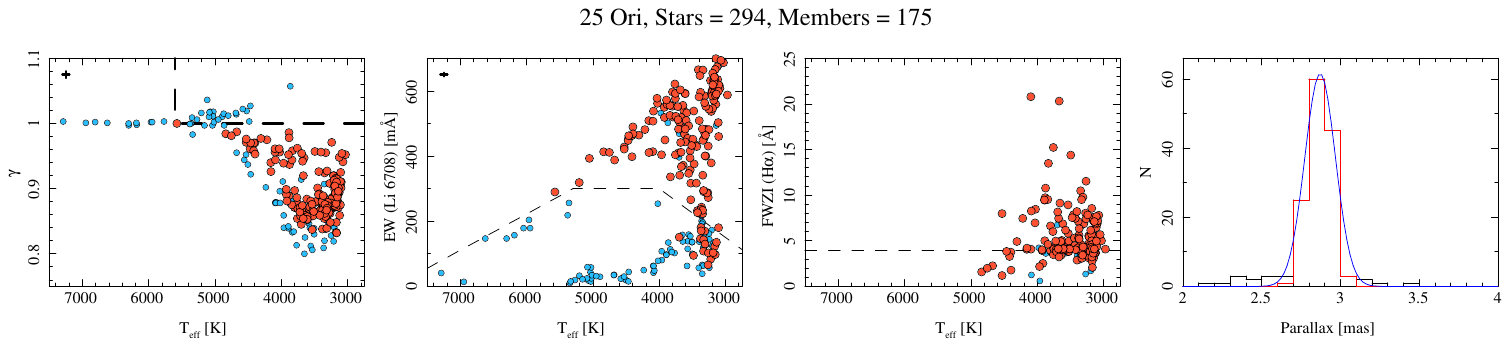}
\includegraphics[width=14.5cm,trim=100 110 100 340]{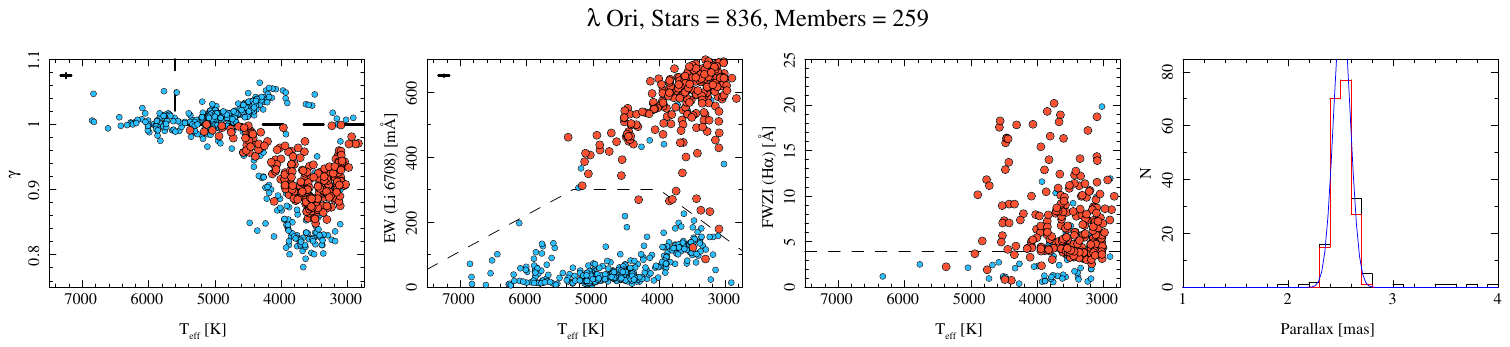}
\caption{Quantities derived from GES spectroscopy used for membership selection and parallax distributions for all groups. Left-hand panels show the gravity index $\gamma$, centre-left panels show the EW of the lithium 6708~\AA\ line, and centre-right-hand panels show the FWZI of H$\alpha$, all plotted against effective temperature. In all three panels red circles show sources that pass our spectroscopic membership criteria and blue circles show sources that do not. The dashed lines show the thresholds used to identify giants in the upper-right corner of the $\gamma$-$T_{eff}$ plot and to identify young stars above the thresholds in the EW(Li)-$T_{eff}$ and FWZI(H$\alpha$)-$T_{eff}$ plots. A typical error bar is shown in the top-left corner of each panel illustrating the typical uncertainties of 100~K in $T_{eff}$, 13~m\AA\ for EW(Li), 0.011 for $\gamma$ and 1.1~\AA for FWZI(H$\alpha$). The right-hand panel shows the parallax distribution for the likely members towards each group. The black histogram shows the distribution of spectroscopically-identified members, the blue line shows a Gaussian fit to this distribution, and the red histogram shows the distribution of the sources that fall within 2 standard deviations (also accounting for errors) of the median parallax and therefore that constitute our final sample.}
\label{spectral_membership1}
\end{figure*}

\begin{figure*}
\includegraphics[width=14.5cm,trim=100 110 100 340]{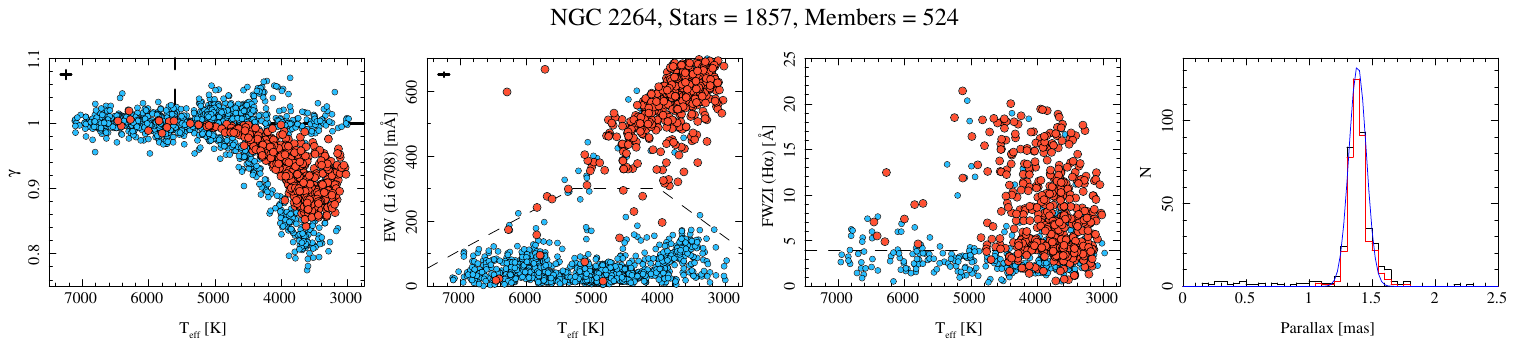}
\includegraphics[width=14.5cm,trim=100 110 100 340]{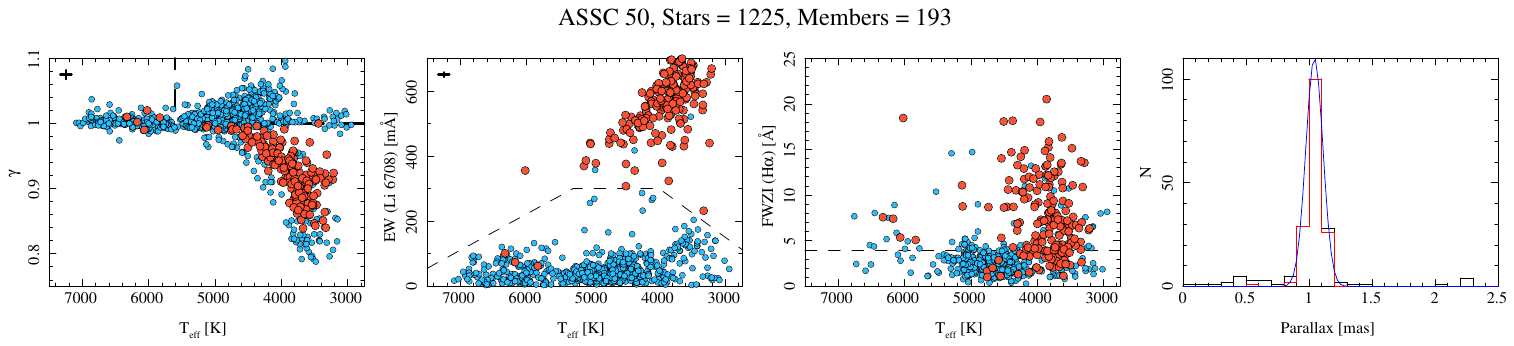}
\includegraphics[width=14.5cm,trim=100 110 100 340]{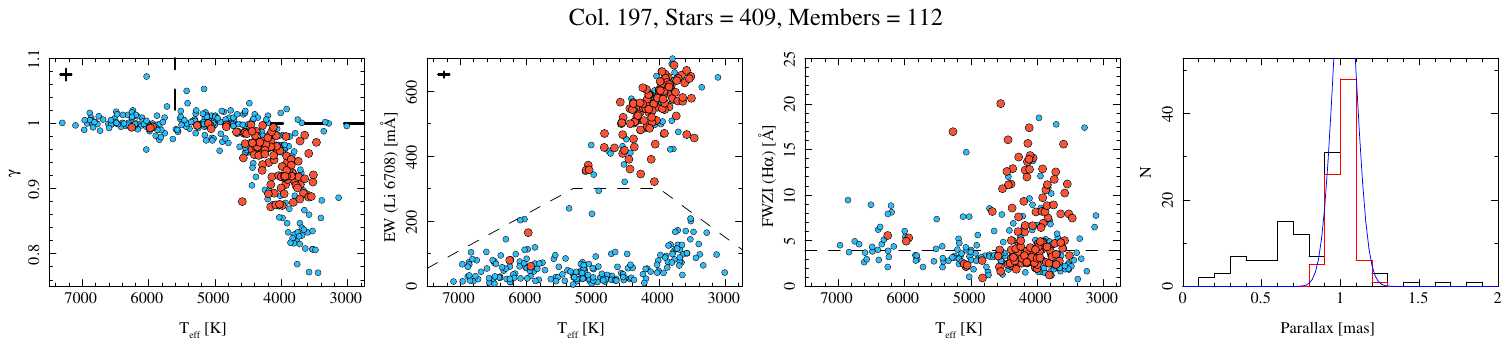}
\includegraphics[width=14.5cm,trim=100 110 100 340]{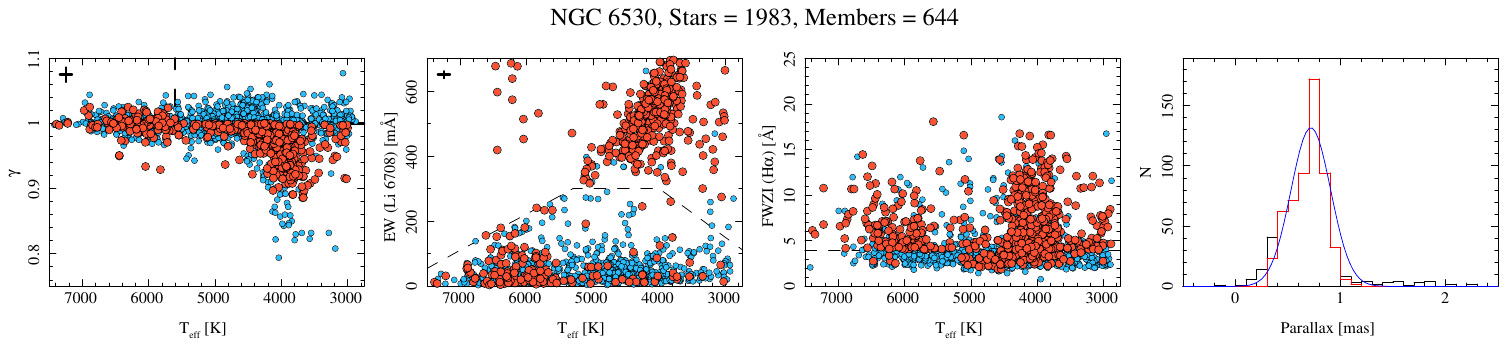}
\includegraphics[width=14.5cm,trim=100 110 100 340]{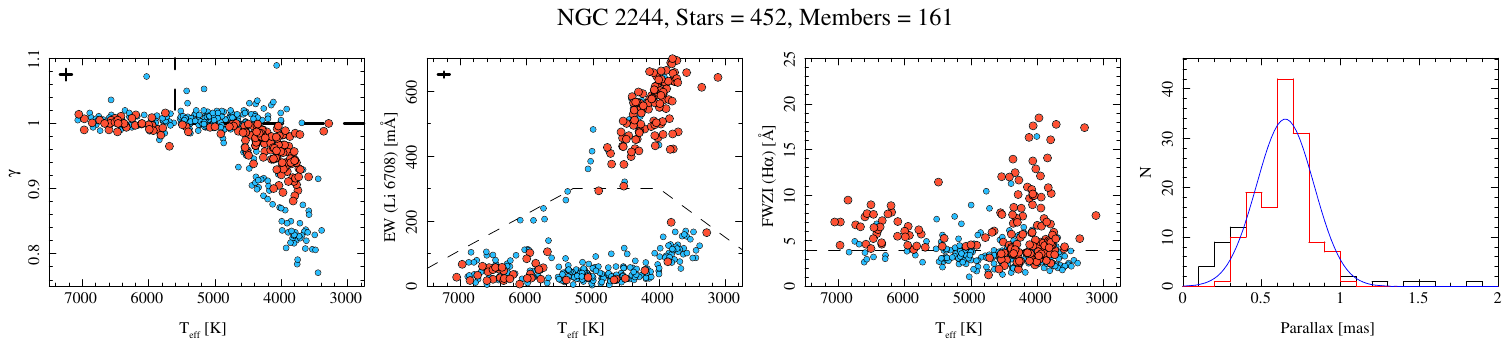}
\includegraphics[width=14.5cm,trim=100 110 100 340]{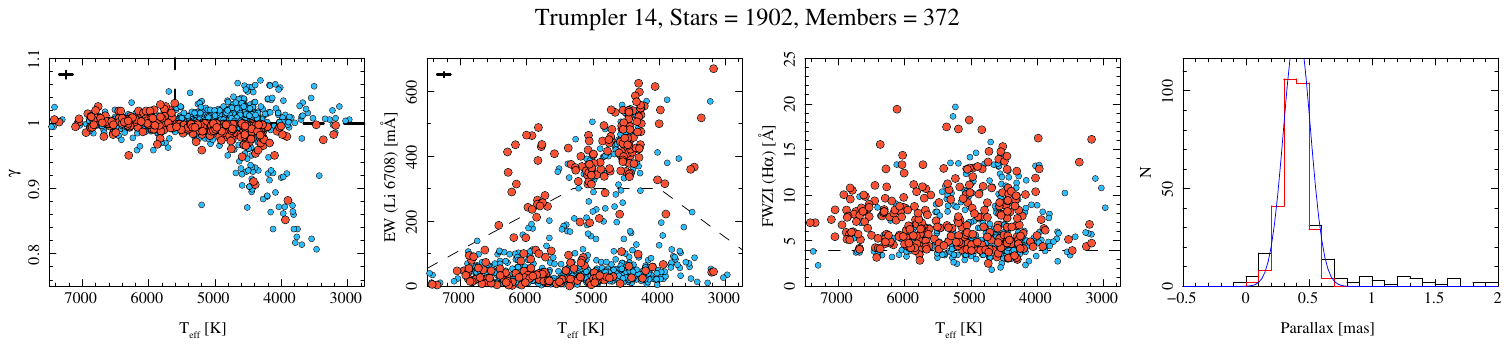}
\caption{As per Figure~\ref{spectral_membership1}.}
\label{spectral_membership2}
\end{figure*}

Here we provide a summary of our young star membership selection process, including figures illustrating the process in Figures \ref{spectral_membership1} and \ref{spectral_membership2}. For a star to be included as a young star in our catalogue it must pass the following tests:

\begin{enumerate}
\item The star must have either a lithium equivalent width greater than observed in stars of the same temperature in the 30-50 Myr cluster IC 2602 \citep{rand97} or they must have H$\alpha$ full width at zero intensity (FWZI) measurements greater than 4 \AA \citep{boni13}.
\item The star must not have a gravity-sensitive index $\gamma > 1$ and $T_{eff} < 5600$~K \citep[which together indicate the star is a cool giant,][]{dami14}.
\item The star must have a parallax within 2$\sigma$ of the central value determined for the group from Gaussian fitting to the parallax dispersion.
\end {enumerate}

Stars are not required to have valid RVs or PMs to be included in our overall sample of 2683 young stars, and therefore stars whose astrometry does not pass the {\it Gaia} RUWE $\leq 1.4$ quality cut have their astrometry discarded but they themselves, and their RVs, are not discarded. This decision was made to maximise the number of stars with reliable kinematic data in at least one dimension that we can use to constrain the kinematic properties of the groups and clusters studied. Our kinematic analysis is of course limited to stars with at least an RV or a PM.

\section{Proper motion vector point diagrams}

\begin{figure*}
\includegraphics{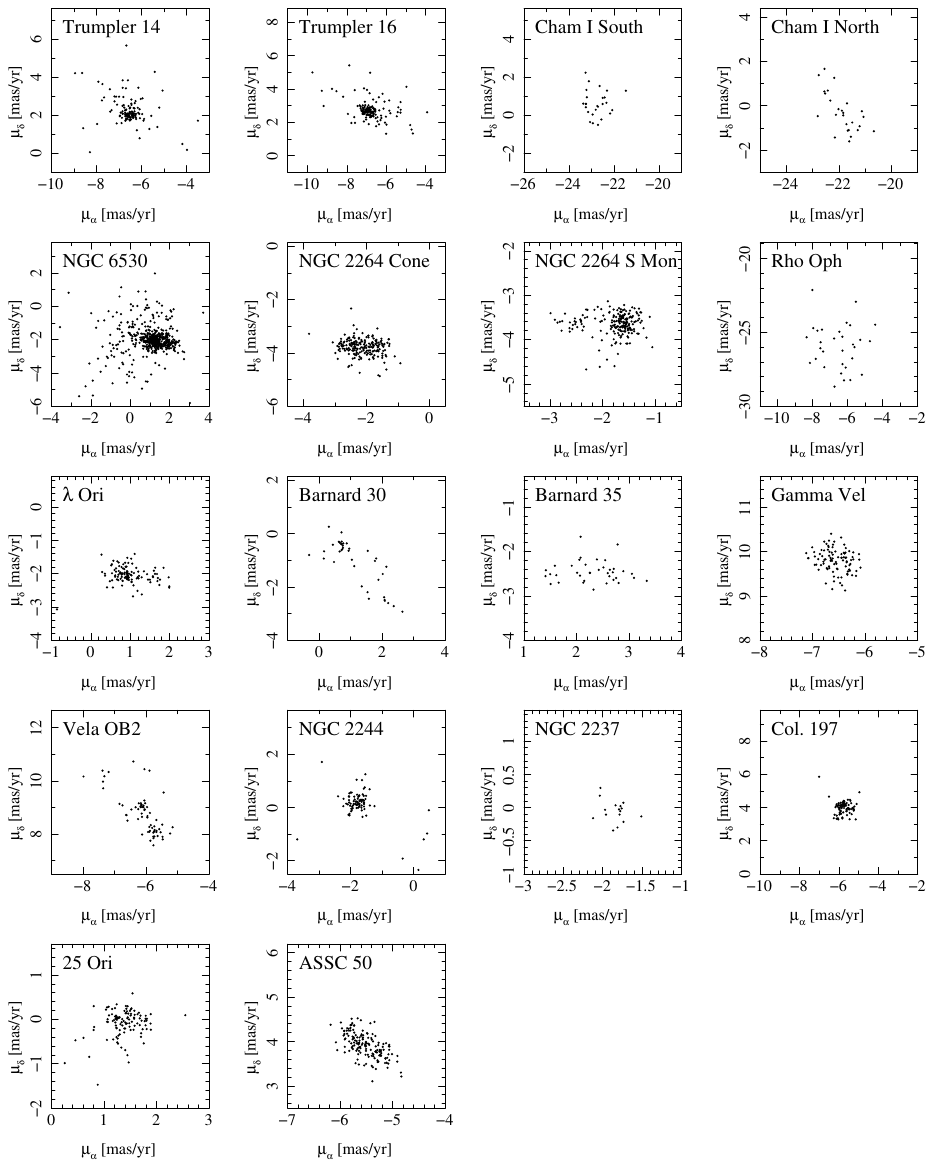}
\caption{PM ``vector point" diagrams for the members of all groups with available proper motions. A reminder that the median PM uncertainty is 0.062 mas~yr$^{-1}$, which in the vast majority of cases is smaller than the symbol size used.}
\label{vector_point_diagrams}
\end{figure*}

Here we provide in Figure~\ref{vector_point_diagrams} the proper motion ``vector point'' diagrams for all groups and clusters, showing the members of each group with valid proper motions.

\end{document}